\documentclass[aps,prb,amsmath,amssymb,floatfix,twocolumn]{revtex4-1}
\usepackage{graphicx}
\usepackage{subfigure}
\usepackage{amsmath}
\usepackage{bbold}
\usepackage{slashed}
\usepackage{amssymb}
\usepackage{braket}
\usepackage{array}
\usepackage[colorlinks]{hyperref}
\usepackage{float}
\usepackage[margin=1in]{geometry}

\newcommand{\ncmd}{\newcommand}
\ncmd{\nn}{\nonumber}
\ncmd{\pg}[1]{\textcolor{red}{#1}}
\ncmd{\mbf}[1]{\bs{#1}}
\ncmd{\Lam}{\Lambda}
\ncmd{\lam}{\lambda}
\ncmd{\Gam}{\Gamma}
\ncmd{\gam}{\gamma}
\ncmd{\sig}{\sigma}
\ncmd{\Dl}{\Delta}
\ncmd{\dl}{\delta}
\ncmd{\kap}{\kappa}
\ncmd{\Om}{\Omega}
\ncmd{\om}{\omega}
\ncmd{\mc}{\mathcal}
\ncmd{\eps}{\epsilon}
\ncmd{\veps}{\varepsilon}
\ncmd{\vphi}{\varphi}
\ncmd{\vtheta}{\vartheta}
\ncmd{\note}[1]{{\color{red}{#1}}}
\ncmd{\new}[1]{{\texttt{#1}  } }
\ncmd{\eq}[1]{Eq. \eqref{#1}}
\ncmd{\bs}{\boldsymbol}
\ncmd{\pll}{\parallel}
\ncmd{\dsty}{\displaystyle}

\begin{document}

\title{Non-Abelian Stokes theorem and quantized Berry flux}
\author{Alexander C. Tyner$^{1}$, Shouvik Sur$^{2}$, Qunfei Zhou$^{3,4}$, Danilo Puggioni$^{5}$, Pierre Darancet$^{4,6}$, James M. Rondinelli$^{1,5,6}$, and Pallab Goswami$^{1,2}$}
\affiliation{$^{1}$ Graduate Program in Applied Physics, Northwestern University, Evanston, Illinois, 60208, USA}
\affiliation{$^{2}$ Department of Physics and Astronomy, Northwestern University, Evanston, Illinois, 60208, USA}
\affiliation{$^{3}$ Materials Research Science and Engineering Center, Northwestern University, Evanston, IL, 60208, USA}
\affiliation{$^{4}$ Center for Nanoscale Materials, Argonne National Laboratory, Argonne, IL, 60439, USA}
\affiliation{$^{5}$ Department of Materials Science and Engineering, Northwestern University, Evanston, Illinois, 60208, USA}
\affiliation{$^{6}$ Northwestern Argonne Institute for Science and Engineering, Evanston, IL, 60208, USA}

\date{\today}

\begin{abstract}
Band topology of anomalous quantum Hall insulators can be precisely addressed by computing Chern numbers of constituent non-degenerate bands that describe quantized, Abelian Berry flux through two-dimensional Brillouin zone. Can Chern numbers be defined for $SU(2)$ Berry connection of two-fold degenerate bands of materials preserving space-inversion ($\mathcal{P}$) and time-reversal ($\mathcal{T}$) symmetries or combined $\mathcal{PT}$ symmetry, without detailed knowledge of underlying basis? We affirmatively answer this question by employing a non-Abelian generalization of Stokes' theorem and describe a manifestly gauge-invariant method for computing magnitudes of quantized $SU(2)$ Berry flux (spin-Chern number) from eigenvalues of Wilson loops. The power of this method is elucidated by performing $\mathbb{N}$-classification of \emph{ab initio} band structures of three-dimensional, Dirac materials. Our work outlines a unified framework for addressing first-order and higher-order topology of insulators and semimetals, without relying on detailed symmetry data.
\end{abstract}

\maketitle

\section{Introduction}Niu1985,
The basic concepts of topological band theory were developed by considering global properties of non-degenerate energy bands of time-reversal symmetry breaking, two-dimensional insulators~\cite{TKNN1982,PhysRevLett.61.2015, Niu1985,Kane2005,Fukui2005,bernevig2006quantum,FuKane,FuKaneMele2007,Moore2007}. The band eigenfunctions of such systems are determined up to arbitrary complex phase factors, i.e. $\psi_{n}(\mathbf{k})$ and $e^{i\alpha_{n}(\mathbf{k})}\psi_{n}(\mathbf{k})$ are equally good candidate eigenfunctions for the $n$-th band, with $n=1,2,3...,N$. This $U(1)$ redundancy for individual bands leads to Abelian Berry's connection, $\mathbf{A}_{n}(\mathbf{k})=-i\bra{\psi_{n}}\mathbf{\nabla}\ket{\psi_{n}}$ and corresponding Berry's curvature $\mathbf{\Omega}_{n}(\mathbf{k})=\mathbf{\nabla} \times \mathbf{A}_{n}(\mathbf{k})$. By integrating $\mathbf{\Omega}^{j}_{n}$ over the two-dimensional Brillouin zone (BZ), one arrives at the quantized flux of $U(1)$ curvature, $\int d^{2}k \mathbf{\Omega}_{n}(\mathbf{k})=2\pi C_{n}$, where $C_{n}$ is the Chern number of band $n$. There exist many reliable methods for computing $C_{n}$. For example, by measuring the Berry's phase accrued by $\psi_{n}(\mathbf{k})$ when it is parallel transported along any non-intersecting closed contour and relating it to enclosed flux by Stokes theorem.

\par 
Can quantized flux exist for two-fold degenerate bands of parity+time-reversal ($\mathcal{PT}$) invariant systems? The two-fold degeneracy gives rise to local $SU(2)$ redundancy of each band; as $\{\psi_{n,\uparrow}(\mathbf{k}),\psi_{n,\downarrow}(\mathbf{k})\}^{T}$, $g_{n}(\mathbf{k})\{\psi_{n,\uparrow}(\mathbf{k}),\psi_{n,\downarrow}(\mathbf{k})\}^{T}$ are equally good candidate wavefunctions. The curvature, $F_{n,s,s'}$, of $SU(2)$ Berry's connections, $\mathbf{A}_{n,s,s'}(\mathbf{k})=-i\bra{\psi_{n,s}(\mathbf{k})}\mathbf{\nabla}\ket{\psi_{n,s}(\mathbf{k})}$, is gauge covariant. Therefore, there are many conceptual subtleties in assigning gauge-invariant non-Abelian Berry's flux. If global symmetries such as $U(1)$ spin-conservation\cite{bernevig2006quantum} or mirror symmetry\cite{TeoMirror} are present, it becomes possible to assign a global spin quantization axis, ($U(1)$ gauge-fixing) of $SU(2)$ connections. In the case of a mirror symmetry, one can separate the eigenspace into two subspaces, labeled by the eigenvalues of the mirror operator, and calculate $C_{n}$ in each subspace. In the absence of mirror symmetry there is currently no suitable, gauge-invariant method, for computation of flux. Topological classification of general $\mathcal{PT}$ symmetric, two-dimensional insulators, preserving $\mathcal{P}$ and $\mathcal{T}$ individually, thus relies on assignment of the Fu-Kane $Z_{2}$ index \cite{FuKane}. Furthermore, generic two dimensional planes, embedded in a three-dimensional system, need not support their own time-reversal invariant momenta (TRIM) points, precluding assignment of even the $Z_{2}$ index. Nevertheless, many such planes have been proposed as forms of higher-order topological insulators\cite{benalcazar2017,schindler2018,MaoLin2018,wieder2020strong}. \emph{Does assignment of flux in a Kramers degenerate band structure require mirror symmetry? Does a $Z_{2}$ index imply the presence of an underlying quantized flux?} In this work we answer these questions utilizing the method of Wilson loops (WLs), demonstrating that quantized flux can be computed for any $n$-fold rotationally symmetric plane through analysis of the non-Abelian Berry's gauge connections\cite{Wilson,THOOFT1979141}. This method is shown to be applicable in both tight-binding models as well as \emph{ab initio} data.

\par
Consider a closed, non-intersecting path lying in the $xy$ plane and respecting the $m$-fold symmetry of the plane. The WL of $SU(2)$ connections of $n$-th Kramers-degenerate bands along this contour, parameterized by $\bs{k}(l)$, is defined as 
\begin{eqnarray}
W_n &=& \; P \exp \left [i \oint \sum_{j=1}^{2} \; A_{j,n} (\bs{k}(l))\; \frac{dk_j }{dl} \; dl \right] , \\
&=& \exp \left[ i \; \theta_{n}(k_{0}) \; \hat{\bs{\Omega}}_n(k_0) \cdot \boldsymbol \sigma \right],
\end{eqnarray} 
where $P$ denotes path ordering and $k_0$ corresponds to the edge size of the loop. The intra-band connections of $n$-th band, are defined according to the formula $A_{j, n, s, s^\prime}(\bs{k})= -i \psi^\dagger_{n,s}(\bs{k}) \partial_j \psi_{n,s^\prime} (\bs{k})$, where $\psi_{n,s} (\bs{k})$ are the eigenfunctions of $n$-th band, with $s=\pm 1$ denoting the Kramers index, and $\partial_j= \frac{\partial}{\partial k_j}$. The gauge invariant angle $\theta_n(k_0)$ can be related to the magnitude of non-Abelian, Berry's flux by employing a non-Abelian generalization of Stokes's theorem \cite{Halpern,Arefeva1980,Bralic,tyner2020topology}. The gauge dependent, three-component, unit vector $\hat{\bs{\Omega}}_n(k_0)$ defining the orientations in $SU(2)$ color space will not be used for computing any physical properties. When the $n$-th Kramers-degenerate bands support quantized flux of magnitude $|2M \pi| $ with $M\in \mathbb{Z}$, $|\Delta \theta_n(k_0)|= |\theta_n(k_0) - \theta_n(0)|$ will interpolate from $0$ to $|2M\pi|$ as $k_0$ is systematically increased from $0$ to a final value $k_f$, when the area enclosed by the loop becomes equal to the area of two-dimensional BZ. Such quantized flux can be found for two-dimensional planes preserving $\mathcal{P}$ symmtery, as well as planes supporting a global $U(1)$ symmetry. In other cases, such as generic two-dimensional insulators which break $\mathcal{P}$ and $\mathcal{T}$ symmetries, but for which the product $\mathcal{PT}$ is preserved, it is possible for non-Abelian flux through the full zone to be non-quantized. This absent quantization in the non-Abelian phase measured by WL in the first BZ indicates the presence of a twisted boundary condition. As will be shown, twisted boundary conditions do not exclude assignment of quantized flux in the extended BZ.  
\par
The current standard for topological analysis of \emph{ab initio} data, involves calculation of Wannier charge centers (WCCs). WCCs, the gauge invariant spectra of straight Wilson loops, also known as Polyakov loops (PL) ~\cite{Wilson} have emerged as powerful tools for describing topology of quasi-particle band-structures~\cite{yu2011,fidkowski2011,soluyanov2011,alexandradinata2014}. Generally PLs are used for diagnosis of the Fu-Kane strong and weak $Z_{2}$ indices and, in the presence of mirror-symmetry or non-degenerate bands, are utilized to determine quantized Berry's flux\cite{Gresch2017}. However, in the absence of mirror-symmetry PLs fail to recognize flux in $\mathcal{PT}$ symmetric band structures. Further, PLs are known to fail in identifying non-trivial topology for the class of higher-order topological insulators (HOTIs). The necessity of the proposed method can be summarized by Fig. \eqref{fig:FluxTable}, detailing the symmetry requirements to assign a two-dimensional bulk invariant based on quantized Berry's flux to a Kramers degenerate band using current methods. The shortfalls of these methods are clear as, in the absence of mirror or $U(1)$ spin conservation symmetry, no existing method can capture the bulk invariant. As a result a large number of systems, particularly those with even integer bulk invariants invisible to the $\mathbb{Z}_{2}$ index, have gone undetected.

\par
We will explicitly demonstrate the power of this method by performing topological classification of \emph{ab initio} band structures of Dirac semimetals (DSMs). We choose to examine DSMs as the generic two-dimensional planes lying between the Dirac nodes and perpendicular to the axis of nodal separation have been identified as examples of two-dimensional higher-order topological insulators with gapped edge states, while the high-symmetry planes are identified by either a $Z_{2}$ index, or a mirror Chern number, depending on the presence of mirror symmetry\cite{benalcazar2017,schindler2018,wieder2020strong}. DSMs thus offer the chance to study distinct types of higher order- and first order- topological insulators, embedded in a single material. 
\par 
Na$_3$Bi was proposed as the first candidate material for realizing stable DSMs, which arise from linear touching between a pair of two-fold, Kramers-degenerate bands at isolated points of momentum space, along an axis of $n$-fold rotation (say the $\hat{z}$ or $c$-axis) \cite{wang2012dirac}. The Dirac points are simultaneously protected by the combination of parity and time-reversal symmetries ($\mathcal{PT}$) and the $n$-fold rotational ($\mathcal{C}_n$) symmetry~\cite{yang2014classification,armitage2018weyl}. The low energy Hamiltonian is written as, $H(\bs{k})=\epsilon_0(\bs{k}) \mathbb{1} + \sum_{j=1}^{5} \; d_j(\bs{k}) \Gamma_j$, where $\Gamma_j$'s are again five, mutually anti-commuting, $4 \times 4$ matrices, and $\mathbb{1}$ is the $4 \times 4$ identity matrix \cite{wang2012dirac}. The topological properties of conduction and valence bands are controlled by the $O(5)$ vector field $d_1=A k_x$, $d_2= A k_y$, $d_3=B k_z(k^2_x-k^2_y)$, $d_4=2B k_x k_y k_z$, and $d_5= M_0 - M_1 k^2_z -M_2 (k^2_x+ k^2_y)$, where $A$, $B$, $M_0$, $M_1$, and $M_2$ are band parameters. For Na$_3$Bi, the parameters $M_0<0$, $M_1<0$, and $M_2<0$ capture band inversion effects, leading to two Dirac points along the six-fold, screw axis at $(0,0, \pm k_D)$, with $k_D=\sqrt{M_0/M_1}$. The particle-hole anisotropy term $\epsilon_0(\bs{k})$ does not affect band topology. 

For describing low-energy physics of massless Dirac fermions, $d_3$ and $d_4$ terms can be ignored in the renormalization group sense \cite{wang2012dirac,Gorbar2015,Burkov2016}. Such approximate theories predict topologically protected, loci of zero-energy surface-states, also known as the \emph{helical Fermi arcs}, joining the projections of bulk Dirac points on the $(100)$ and the $(010)$ surface- Brillouin zones. Therefore, the spectroscopic detection of \emph{helical Fermi arcs} was often considered to be the smoking gun evidence of bulk topology of DSMs. However, these terms cannot be ignored for addressing topological properties of generic planes and they are responsible for gapping out the helical edge states for all $|k_z| < k_D$ and $k_z \neq 0$ \cite{kargarian2016surface,Bednik2018,le2018dirac}, and giving rise to higher-order topology~\cite{MaoLin2018,wieder2020strong}.

\section{Ab initio band structures} The crystal structure of Na$_{3}$Bi, belongs to the space group P6$_3$/mmc and has the lattice constants $a=b=5.49$ \AA, $c=9.78$ \AA. It consists of two non-equivalent Na sites, denoted by Na(1) and Na(2). The honeycomb layers formed by Na(1) and Bi are stacked along the c-axis, with Na(2) sites located between the layers. For computational details please see the supplementary materials.
The calculated band structures within the energy window $-3$ eV and $+2 $ eV are displayed in Fig.~\ref{fig:1b}. We have labeled the Kramers-degenerate bands, according to their energy eigenvalues at the $\Gamma$ point, with $E_{n}(0)<E_{n+1}(0)$. The bulk Dirac points arise from linear touching between bands $n=6$ and $n=7$, along the six-fold, screw axis ($A-\Gamma-A$ line or the $k_{z}$ axis) at $(0,0, \pm k_{D})$, with $k_D \approx \pm 0.29 \times \frac{\pi}{c}$. Their reference energy coincides with the Fermi level.

\subsection{Bulk Topology}
In order to perform topological analysis of various bands, we have employed maximally localized Wannier functions calculated using the WANNIER90 package~\cite{Pizzi2020wannier90}. We will calculate WLs of individual $SU(2)$ Berry's connections of bands $n=1$ through $n=6$ by utilizing the Z2Pack software package \cite{soluyanov2011,Gresch2017}. In calculating WLs, we have followed the hexagonal path $abcdef$, shown in Fig.~\ref{fig:3a}.
\par 
We first focus on the results of the WL for occupied bands in the $k_{z}=0$ mirror plane. In this plane, we find that the Dirac band ($n=6$) as well as four remote bands ($n=1,2,3,5$) support non-zero flux of varying magnitude (see Fig.~\ref{fig:3c}) for which quantization occurs for a contour exactly enclosing the two-dimensional BZ. Due to the presence of mirror symmetry, flux could also have been computed via WCCs, yielding identical results. By contrast, away from the mirror planes, WCCs can no longer be utilized to determine flux. For more details of WCCs in these planes please see the supplementary materials. Computing WLs for the occupied Dirac band at generic topologically non-trivial planes defined by $|k_z| < k_D$, $|\Delta \theta_{6}(k_0)|$ shows $0$ to $2\pi$ interpolation for a contour enclosing an area slightly greater than the two-dimensional BZ. This behavior is in accordance with the low energy lattice model of Na$_{3}$Bi presented in the supplementary information. For $|k_z| > k_D$, $|\Delta \theta_{6}(k_0)|$ does not exhibit such interpolation, indicating that these planes are topologically trivial. These topological properties of Dirac bands are identical to what have been found from the effective, four-band model of $sp$-hybridized DSMs \cite{tyner2020topology}.
\par
While topological diagnosis of the high-symmetry planes of Na$_{3}$Bi benefits from the presence of mirror symmetry, one can also consider a DSM where the high-symmetry plane lying perpendicular to the direction of nodal separation lacks mirror symmetry. One such system is $\beta$-CuI, which was proposed as a DSM by Le et. al\cite{le2018dirac}, with the Dirac nodes lying along the $k_{z}$ axis. For full computational details of this material please consult the supplementary materials. As $\beta$-CuI belongs to space group $R\bar{3}m$, the high-symmetry $xy$ planes support three-fold rotational symmetry. Mirror symmetry is therefore absent and the current topological classification of the planes is limited to assignment of a $Z_{2}$ index. In Fig. \eqref{fig:CuIPWl}, we show that for the high-symmetry and generic planes lying between the Dirac nodes, the method of WLs can be utilized to identify a quantized flux. These results emphasize the presence of flux as the common topological invariant of two-dimensional topological insulators, regardless of the underlying symmetry. 

\subsection{Conclusions} 
Recently, Tyner et. al\cite{tyner2020topology} demonstrated that unlike Weyl semimetals,  Dirac semimetals do not support Fermi arcs, loci of two, degenerate zero energy state beginning and terminating at the projection of the bulk nodes \cite{tyner2020topology}. Rather, a generic plane lying between the Dirac nodes supporting a non-zero relative Chern number supports two, non-degenerate gapped surface states. Only at the high-symmetry mirror planes can we locate gapless points on the surface with the number of gapless points being determined by the magnitude of the mirror Chern number. Using the iterative Greens function method \cite{sancho1985highly} and the Wannnier Tools software package \cite{WU2017}, we plot the spectral density on the (100) surface of Na$_{3}$Bi in Fig. \eqref{fig:Pan_4}. These results verify that at a generic value of $|k_{z}|<k_{D}$, the (100) surface of Na$_{3}$Bi supports gapped surface states. Only at the $k_{z}=0$ mirror plane do we find a gapless state in correspondence with the admission of mirror Chern number $|\mathcal{C}_{m}|=1$, by this plane. 
\par
In recent works, the bulk-boundary correspondence of two dimensional higher order insulators has been characterized by the presence of corner localized states in those corners of a two-dimensional slab that align with the corners of the primitive two-dimensional unit cell\cite{benalcazar2017,schindler2018,wieder2020strong}. Fixing $|k_{z}|<k_{D}$, and performing an exact diagonalization calculation for a slab consisting of $10\times 10$ primitive unit cells in the $xy$ plane, we schematically depict the wavefunction localization of the four corner-localized states present at half filling in Fig. \eqref{fig:CorStates}. The localization pattern of these states captures the bulk-boundary correspondence and verifies the HOTI classification of the $xy$ planes between the Dirac nodes. However, we emphasize that these states are not mid-gap states, well separated from the bulk states\cite{wieder2020strong}, posing a significant challenge for experimental detection and re-enforcing the need for a robust method of bulk classification. 
\par 
In summary, we have proposed a method capable of quantifying non-Abelian flux through any plane regardless of underlying crystal symmetries. We have successfully applied this method to \emph{ab initio} data of multiple Dirac materials. Our results are insensitive to the number of underlying bands, suggesting the topology of real materials can be comprehensively addressed with stable, bulk invariants.

\acknowledgements {A. C. T., S. S., Q. Z., P. D. and P. G. were supported by the National Science Foundation MRSEC program (DMR-1720139) at the Materials Research Center of Northwestern University. D.P. and J.M.R. acknowledge the Army Research Office under Grant No. W911NF-15-1-0017 for financial support and the DOD-HPCMP for computational resources. Use of the Center for Nanoscale Materials (CNM), an Office of Science user facility, was supported by the U.S. Department of Energy, Office of Science, Office of Basic Energy Sciences, under Contract No. DE-AC02-06CH11357. }

\appendix 
\section{In-plane Wilson Loop}  
For concreteness, let us consider the Hamiltonian $H=\sum_{\bs{k}} \Psi^\dagger(\bs{k}) \hat{H}(\bs{k}) \Psi(\bs{k})$, where $\Psi(\bs{k})$ is a four-component spinor, and the Bloch Hamiltonian operator can be written as
\begin{equation}
	\hat{H}(\bs{k})= N_0(\bs k) \mathbb{1} + \sum_{j=1}^{5} \; N_j(\bs{k}) \Gamma_j,
	\label{eq:H4BD}
\end{equation} 
where the $O(5)$ vector field $\bs{N}(\bs{k})$ encodes details of band-structures,  and $\Gamma_j$'s are five, mutually anti-commuting, $4 \times 4$ matrices, such that $\{\Gamma_i, \Gamma_j\}=2\delta_{ij}$. Specifically we set, $\Gamma_{i=1,2,3}=\tau_{i}\otimes \sigma_{1},\; \Gamma_{4}=\tau_{2}\otimes \sigma_{0}, \Gamma_{5}=\tau_{3} \otimes \sigma_{0}$, where $\tau_{0}$ and $\sigma_{0}$ are two $2 \times 2$ identity matrices. The two sets of Pauli matrices $\tau_{j}$ and $\sigma_{j}$ with $j=1,2,3$ operate on the spin and orbital degrees of freedom respectively. For clarity, we consider 
\begin{align}
&\bs{N}(\bs{k})=[t_p \sin k_x, t_p\sin k_y, t_{d}\sin k_{z}  (\cos k_x -  \cos k_y),  \nonumber \\ 
& t_{d}\sin k_z \sin k_x \sin k_y,  t_s(\Delta - \cos k_x - \cos k_y - \cos k_z)],
\label{O(5)}
\end{align}
which describes $\mc C_4$-symmetric DSMs. 
Here, $t_s$, $t_p$, $t_{d}$ are independent hopping functions.
The dimensionless parameter $\Delta$ controls topological phase transitions. 
In particular, when $1<\Delta<3$, all five components of $\bs{N}(\bs{k})$ vanish at the Dirac points, located at $\bs{k}=(0,0,\pm k_{D,1})$ with $\cos k_{D,1}=(\Delta-2)$. 
\par
It is convenient to compute WL by following a $\mc C_4$ symmetric path, denoted $ABCD$. An image of this path is available in the supplementary information. Without any loss of generality we will choose the point A with $(k_x,k_y)=(-k_0, -k_0)$ as our reference point. When the Kramers-degenerate wave functions are parallel transported between an initial point $\bs{k}_i$ and a final point $\bs{k}_f$, the matrix-valued, non-Abelian Berry's phase.~\cite{demler1999,Zee1984,Wilson} is described by the Wilson line (or non-Abelian holonomy)
\begin{eqnarray}
&&W_{i,f}=\; P \exp [i \int^{l_f}_{l_i} \sum_{j=1}^{2} \; a_j (\bs{k}(l))\; \frac{dk_j }{dl} \; dl] , \label{WL1}
\end{eqnarray}
where $P$ denotes path ordering, and we choose to work with the gauge choice, 
\begin{eqnarray}
&& a_{j}(\bs{k}) =  \frac{1}{2|\bs{N}|(|\bs{N}|+N_5)} [(N_1 \partial_{j} N_{2} - N_2 \partial_{j} N_{1}) \Gamma_{12}  \nonumber \\
&&+(N_2 \partial_{j} N_{3} - N_3 \partial_{j} N_{2}) \Gamma_{23}+(N_3 \partial_{j} N_{1} - N_1 \partial_{j} N_{3}) \Gamma_{31}  \nonumber \\
&& +\sum_{a=1}^{3}(N_a \partial_{j} N_{4} - N_4 \partial_{j} N_{a}) \Gamma_{a4}], \label{singularconnection}\end{eqnarray}
where $\Gamma_{ij}=[\Gamma_{i},\Gamma_{j}]/(2i)$. We note that $a_{j}(\bs{k})$ is singular at TRIM locations in which $N_{5}=-|\bs{N}|$ as discussed in Tyner et. al\cite{tyner2020topology}. We have parameterized the line, joining two points as $k_j(l)$, $\bs{k}_i= \bs{k}(l_i)$ and $\bs{k}_f= \bs{k}(l_f)$. Therefore, the WL for path $ABCD$ can be obtained as the ordered product of four straight Wilson lines as
\begin{eqnarray}
W_{ABCD}(k_0)=W_{A,B}W_{B,C}W_{C,D}W_{D,A}. \label{WL5}
\end{eqnarray}  
Since $W_{ABCD}(k_0) \in Spin(4)$, we can parametrize it as
\begin{eqnarray}
&&W_{ABCD}(k_0)=W_{ABCD,c}(k_0) W_{ABCD, v}(k_0) \nonumber \\
&&=\begin{bmatrix}
\exp [i \theta_{c}(k_0) \;  \hat{\bs{n}}_{cj}(k_0) \cdot \boldsymbol \sigma] & 0 \\
0 & \exp [i \theta_{v}(k_0) \;  \hat{\bs{n}}_{v}(k_0) \cdot \boldsymbol \sigma] \\
\end{bmatrix} .\label{WL6} \nn \\
\end{eqnarray}
Here, $W_{ABCD,c}(k_0) \in SU(2)$ and $W_{ABCD,v}(k_0) \in SU(2)$ are the WLs for the respective $SU(2)$ connections of conduction and valence bands. Two angles $\theta_{c}(k_0)$ and $\theta_{v}(k_0)$ are gauge-invariant and two $O(3)$ unit vectors $\hat{\bs{n}}_{c}(k_0) $ and $\hat{\bs{n}}_{v}(k_0)$ define gauge-dependent orientations in color space. 
\par
If we wish to abstain from making a gauge choice and compute WLs in a purely numerical fashion, the straight Wilson line along $\hat{j}$ for band $n$ can be rewritten as \cite{yu2011,alexandradinata2014,benalcazar2017}, 
\begin{equation}\label{eq:ProjWL}
    W_{j}(\mathbf{k})=F_{j,\mathbf{k}+N_{j}\Delta k_{j}}...F_{j,\mathbf{k}+\Delta k_{j}}F_{j,\mathbf{k}},
\end{equation}
where $F_{j,\mathbf{k}+N_{j}\Delta k_{j}}=\bra{u^{n}_{\mathbf{k}+\Delta k_{j}}}\ket{u^{n}_{\mathbf{k}}}$, $\Delta k_{j}=2\pi/N_{j}$, and $\ket{u^{n}_{\mathbf{k}}}$ is the Bloch function of band $n$ at $\mathbf{k}$. 
\par 
From $\theta_{c/v}(k_0)$ we can construct other gauge-invariant quantities, such as the eigenvalues of WLs $\exp [ \pm i \theta_{c/v}(k_0)] $, the trace of WLs $\mathrm{Tr}[W_{ABCD,c/v}](k_0)=2 \cos[\theta_{c/v}(k_0)]$, and the Vandermonde determinant $D_V[W_{ABCD,c/v}](k_0) = 2i \sin [\theta_{c/v}(k_0)]$. In gauge theory literature, $\mathrm{Tr}[W_{C}]$ is the most widely studied observable. It is useful for detecting interpolation of $W_{C}$ between the center elements $\pm \sigma_0$ of $SU(2)$ group, leading to $\mathrm{Tr}[W_{C}] = \pm 2$. When $\theta_{c,v}=2 l \pi$ [$(2l+1)\pi)$], with $l \in \mathbb{Z}$, $W_{C,c/v}=\sigma_0 $ [$-\sigma_0$]. We determine both $\mathrm{Tr}[W_{ABCD,c/v}]$ and $D_V[W_{ABCD,c/v}]$ to find $\theta_{c/v}$. 
\par
For Abelian connections of non-degenerate bands, the Stokes's theorem directly relates $\theta_c$ and $\theta_v$ to the underlying Berry's flux. In the non-degenerate case, as $k_0$ is tuned from $0$ to $\pi$, $\theta_{c/v}(k_0)$ can interpolate between $0$ and $2 l \pi$, $l \in \mathbb{Z}$. The Chern number is precisely given by $\theta_{c/v}(k_0=\pi)$. The windings of $\theta_{c}$ and $\theta_{v}$ for the non-Abelian connections also indicate the presence of chromo-magnetic flux, and non-trivial second homotopy classification. However, the interpretation of flux requires a non-Abelian generalization of Stokes's theorem~\cite{Halpern,Bralic,Arefeva1980,Fishbane1981,Diakonov,Kondo}, and $\theta_{c,v}$ can be related to the surface-ordered, integrals of parallel-transported, non-Abelian curvatures
\begin{eqnarray}
W_{xy} &=& \; P_s \exp \bigg[i \int d^2k \; W^\dagger_{A,O}(k_x,k_y) f_{xy}(k_x,k_y) \nonumber \\ && W_{A,O}(k_x,k_y) \bigg],\label{NAST}
\end{eqnarray}
where, $P_s$ denotes surface ordering, and $f_{xy}(k_x,k_y)=\partial_x a_y - \partial_y a_x + i [a_x, a_y]$ corresponds to covariant curvatures. Here, $W_{A,O}(k_x,k_y)$ is the parallel transport operator, defined in Eq.~(\ref{WL1}), whose initial and final points are respectively located at $\bs{k}=(-k_0, -k_0)$ and $\bs{k}=(k_x, k_y)$, shown schematically in Fig. \eqref{fig:PWL}. 

\par

For the model given by eq. \eqref{O(5)}, all topologically non-trivial, $xy$ planes with $|k_z|<k_D$, $\theta_{c}(k_0)$ and $\theta_v(k_0)$ display non-trivial windings, as the size of loop $ABCD$ is systematically increased. For the parameters chosen, ($t_{p}/t_{s}=0.75$ and  $t_{d}/t_{s}=0.5$), convergence to the quantized result is found for a contour only slightly larger than the entire two-dimensional, BZ torus. This causes $W_{ABCD}$ to interpolate between $\mathbb{Z}_2$ centers of the respective $SU(2)$ subgroups. For some intermediate value of $0<k_0<\pi$, $\theta_{c/v}$ reaches $\pi$. Topologically trivial planes do not show such interpolations. On the other hand, generic planes lying between quadratically dispersing Dirac nodes display $0$ to $4\pi$ windings. Since this method does not require any detailed knowledge of the underlying basis, it can be efficiently used for diagnosing bulk topology from \emph{ab initio} band-structures. In contrast to the Abelian projections \cite{tyner2020topology}, this method can only detect the absolute magnitude of flux.

\subsection{Analytical results}
\par 
In the above description of the Wilson loop and in the low-energy derivation presented in the subsequent section, identification of quantized flux relies on tracking interpolation of the WL between $SU(2)$ center elements $\pm \sigma_{0}$. A continuous interpolation between these elements is taken to be in correspondence with the interpolation of the non-Abelian flux, $\theta_{c/v}$, by $2\pi$ as Tr$W_{xy}=2\cos\theta_{c/v}$. Here we provide proof of the correspondence between non-Abelian loop and quantized non-Abelian flux. To do so, we consider the model, 
\begin{eqnarray}\label{eq:naBHZ}
     H(\mathbf{k})&=&\sin \alpha(k_{\perp}) \cos(m\phi)\Gamma_{1}+\sin \alpha(k_{\perp}) \sin(m\phi)\Gamma_{2} \nn \\ && +\cos \alpha(k_{\perp})\Gamma_{3}
\end{eqnarray}
where $\Gamma_{i}$'s are defined previously and $\cos \alpha(k_{\perp})=(\Lambda-k_{\perp}^{2m})/(\Lambda+k_{\perp}^{2m})$, with $\Lambda$ defining the skyrmion core size. This model is chosen to prove correspondence with non-Abelian flux as all WLs are analytically tractable. When $\text{sgn}(\Lambda)>(<)0$, this model is in a topological (trivial) phase for which it is known to support quantized $|2m\pi|$ (0) Berry's flux\cite{bernevig2006quantum}. However, in the current basis, this model is block off-diagonal, and the $SU(2)$ intra-band Berry's gauge connections are matrix valued, i.e. non-Abelian. In order to prove that the non-Abelian Wilson loop precisely measures flux, we consider a path enclosing a sector of the circular Brillouin zone with central angle $2\pi/N$. The non-Abelian WL can thus be written as,
\begin{eqnarray}
    W_{N}&=&W_{k_{\perp}, k_{\perp}=0}^{k_{\perp}=\infty}(\phi=\phi_{i})W_{\phi, \phi_{i}}^{\phi_{i}+2\pi/N}(k_{\perp=\infty}) \nn \\ && W_{k_{\perp}, k_{\perp}=\infty}^{k_{\perp}=0}(\phi=\phi_{i}+2\pi/N).
\end{eqnarray}
After some algebra, using the definition for the intra-band Berry's gauge connections written previously, we arrive at the form, 
\begin{equation}
   \frac{1}{2}  Tr [W_{N}]=\cos^{2} \Omega+\cos (2m\pi/N)\sin^2 \Omega,
\end{equation}
where, 
\begin{equation}
    \Omega=\begin{cases}
    \pi/2, \; \text{sgn}(\Lambda)>0\\
    0, \; \text{sgn}(\Lambda)<0
    \end{cases}.
\end{equation}
Therefore we can conclude that in the topological phase, $\text{sgn}(\Lambda)>0$, Tr$W_{N}/2=\cos (2m\pi/N)$. As such we have directly shown correspondence between quantized Berry's flux, $\theta_{c/v}$ and the WL, validating the above procedure of tracking $SU(2)$ center elements. 
\par 
Additionally, we could consider analyzing $W_{\phi}(\mathbf{k}_{\perp})$ as a function of $k_{\perp}$. After some algebra, we arrive at the expression, 
\begin{equation}
    \text{Tr}W_{\phi}(\mathbf{k}_{\perp})/2=\cos(m\pi)\cos(m \pi \sqrt{10-6\cos(2\alpha)}/2).
\end{equation}
We note that in the simple case, $m=1$, $\text{Tr}W_{\phi}(\mathbf{k}_{\perp})/2=-1$ along the curve defining band inversion, namely $k_{\perp}^{2m}=\Lambda$. However, such correspondence vanishes for $m>1$, as $W_{\phi}(\mathbf{k}_{\perp})=\pm \sigma_{0}$ at general values of $k_{\perp}$, which are not in correspondence with the location of band inversion. 

In order to analytically demonstrate the effectiveness of WLs in detecting the presence of quantized non-Abelian flux for five-component models, such as those describing higher-order insulators, we consider the following continuum version of \eqref{eq:H4BD},
\begin{eqnarray}
	&&\hat{H}(\bs{k})= \sum_{j=1}^{5} \; N_j(\bs{k}) \Gamma_j=B_{1}k_{\perp}(\cos \phi\Gamma_{1}+\sin \phi \Gamma_{2})  \nn \\ &&+B_{2}(k_{z})k_{\perp}^2(\cos 2\phi\Gamma_{3}+\sin 2\phi \Gamma_{4})+N_{5}(k_{\perp},k_{z})\Gamma_{5}, \nn \\
	\label{eq:Hkp}
\end{eqnarray} 
where $(k_{x},k_{y})=k_{\perp}(\cos \phi, \sin \phi)$. We define $N_{5}(k_{\perp},k_{z})$, such that for $k_{\perp}=0(\infty),|k_{z}|<k_{D}$, $N_{5}/\mathbf{N}=-(+)1$ while for $k_{\perp}=0(\infty),|k_{z}|>k_{D}$, $N_{5}/\mathbf{N}=+(+)1$. The non-Abelian gauge connection, $A_{\phi}$, can then be calculated following eq. \eqref{singularconnection}. The Wilson loop follows as, 
\begin{align}
	W^{\pm}_{\phi}(k_{\perp},k_{z})=\mathcal{P}\text{exp}\left[i \int_{0}^{2\pi} A_{\phi}(\mathbf{k}) d_{\phi} \right], 
\end{align} 
where $(\pm)$ indicates the Kramers degenerate conduction and valence bands respectively. Defining $W^{\pm}_{\phi}(k_{\perp},k_{z})=\text{exp}\left(i\Phi^{\pm}\cdot \mathbf{\sigma}\right)$, we examine the gauge invariant quantity $\cos \Phi^{\pm}=-\cos \alpha^{\pm}$. This quantity is equivalent to $\text{Tr}(W^{\pm}_{\phi})/2$. Solving for $\alpha^{\pm}$, we arrive at the form,
\begin{eqnarray}
    \alpha^{\pm}&=&\pi\bigg[\left(\frac{(B_{1}^2\pm2B_{2}(k_{z})^2k_{\perp}^2)k_{\perp}^2}{\mathbf{N}(\mathbf{N}+N_{5})}-1\mp 2\right)^2 \nn \\ &&+\frac{B_{1}^2B_{2}(k_{z})^2k_{\perp}^6}{\mathbf{N}^2(\mathbf{N}+N_{5})^2}\bigg]^\frac{1}{2}.
\end{eqnarray}
We will now investigate this quantity in three important limits (1) at the nodal plane, (2) at the mirror plane, and (3) at a generic plane. 
\par 
\emph{Nodal Plane:}
At the nodal plane, $|k_{z}|=k_{D}$, as $k_{\perp} \rightarrow 0$, $\mathbf{N}$ scales as $k_{\perp}$. As a result $\alpha^{\pm}(k_{\perp\rightarrow 0})=2\pi$ while in the limit $k_{\perp} \rightarrow \infty$, $\alpha^{\pm}(k_{\perp\rightarrow \infty})=\pi|-1\mp2|$. We therefore find the quantized flux in the nodal plane to be $|\Delta\Phi|=|\Phi(k_{\perp}=\infty)-\Phi(k_{\perp}=0)|=\pi$, the critical value. 
\par 
\emph{Mirror Plane:}
For the current model, $B_{2}(k_{z})=k_{z}B_{2}$, therefore at the mirror plane, $k_{z}=0$, we set $B_{2}(k_{z})=0$. In order for a plane to support quantized non-Abelian flux of magnitude $2\pi$, we must be able to show that Tr$W_{\phi}^{\pm}$ evolves adiabatically from $+2\rightarrow  +2$, through Tr$W_{\phi}^{\pm}=-2$. Similar to the calculation of WCCs, this adiabtaic evolution must be treated carefully as it is possible to have a plane in which $|\Delta \Phi|$ exceeds $\pi$ as a function of $k_{\perp}$ before returning to zero. When analyzing Tr$W_{\phi}^{\pm}$, this manifests as two locations of $k_{\perp}$ at which Tr$W_{\phi}^{\pm}=-2$ without reaching Tr$W_{\phi}^{\pm}=+2$ at an intermediate value. We note $\alpha^{\pm}(k_{\perp\rightarrow 0,\infty})=\pi|-1\mp2|$, while the values of $k_{\perp}$ for which Tr$W_{\phi}^{\pm}=-2$ at the mirror plane are found by solving, 
\begin{align}
    |\frac{B_{1}^2k_{\perp}^2}{\mathbf{N}(\mathbf{N}+N_{5})}-1\mp 2|=2n, n\in \mathcal{N}.
\end{align}
This is satisfied if there exists a value of $k_{\perp}$ where $N_{5}=0$, thus the mirror plane supports quantized non-Abelian flux of magnitude $2\pi$. 
\par 
\emph{Generic Plane:}
At a generic value of $k_{z}$, we can again conclude that $\alpha^{\pm}(k_{\perp\rightarrow 0,\infty})=\pi|-1\mp2|$, however we can no longer analytically determine how $\alpha^{\pm}$ interpolates between these values. We thus solve numerically, fixing $B_{2}(k_{z})=k_{z}B_{2}$ and $N_{5}=(B_{3}k_{\perp}^4+B_{4}k_{z}^2-\Delta)$. This process indicates that only planes for which $|k_{z}|<k_{D}$ support quantized non-Abelian flux of magnitude $2\pi$.  

\begin{figure*}[t]
    \centering
    \includegraphics[scale=0.6]{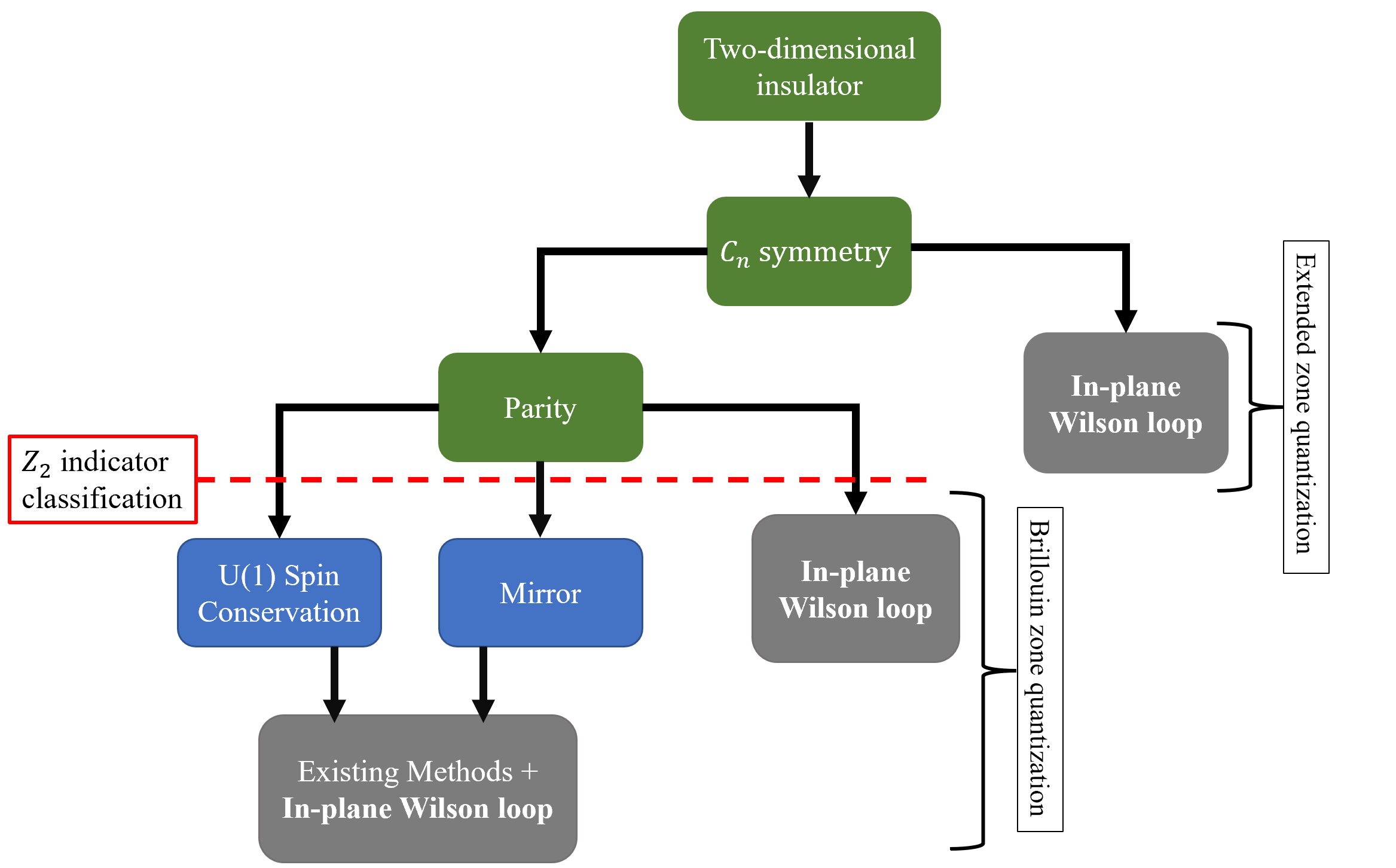}
    \caption{Guide for assignment of bulk invariant based on magnitude of quantized Berry's flux for two-dimensional insulators. All crystalline insulators support $C_{n}$ symmetry. If parity symmetry is additionally present the Fu-Kane $\mathbb{Z}_{2}$ index can be assigned; however this index can not detail the magnitude of the bulk invariant and does not capture systems supporting an even integer bulk invariant. If in addition to parity, mirror or $U(1)$ spin conservation symmetries are present, there are existing methods for obtaining the bulk invariant. However, we emphasize that these symmetries are not common in nature, particularly when considering two dimensional planes, embedded in three dimensions. Nevertheless, the in-plane Wilson loop is capable of capturing quantized flux in each case. If only $C_{n}$ symmetry is present, an extended zone scheme is required.}
    \label{fig:FluxTable}
\end{figure*}

\begin{figure*}[t]
\centering
\subfigure[]{
\includegraphics[scale=0.2]{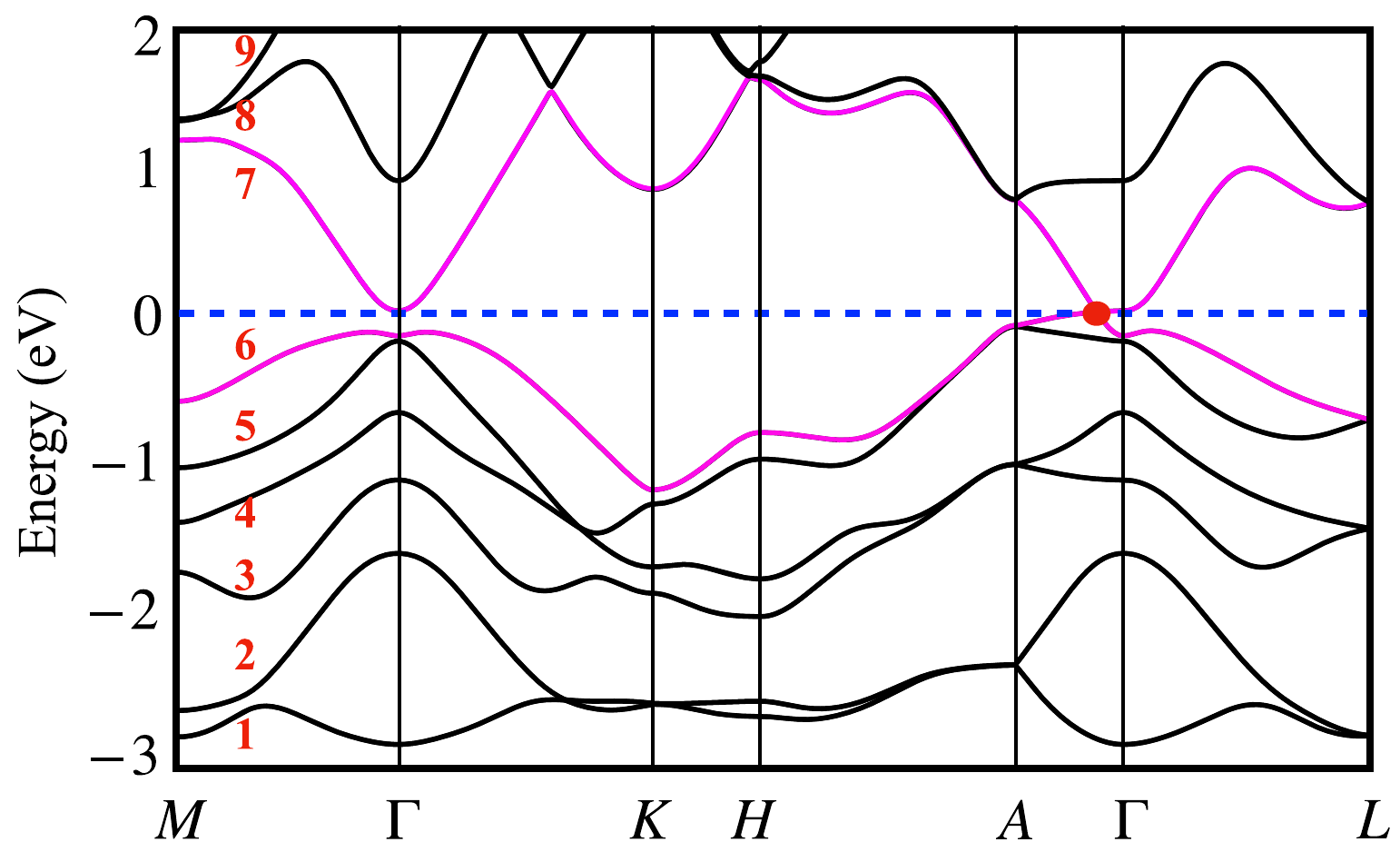}
\label{fig:1b}}
\subfigure[]{
\includegraphics[scale=0.65]{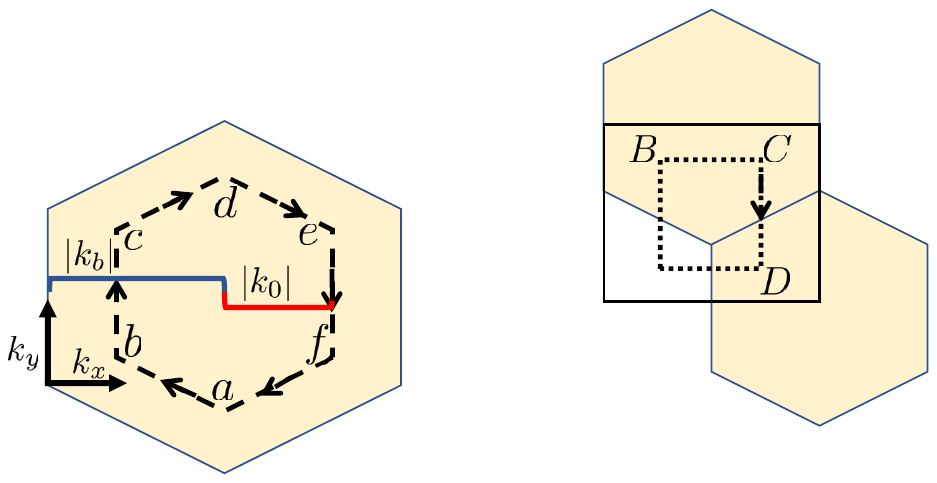}
\label{fig:3a}}
\subfigure[]{
\includegraphics[scale=0.4]{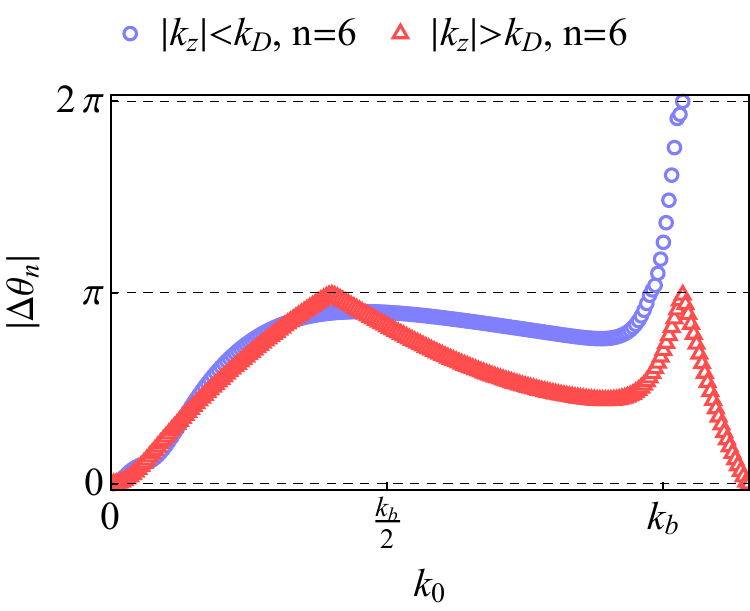}
\label{fig:3b}}
\subfigure[]{
\includegraphics[scale=0.4]{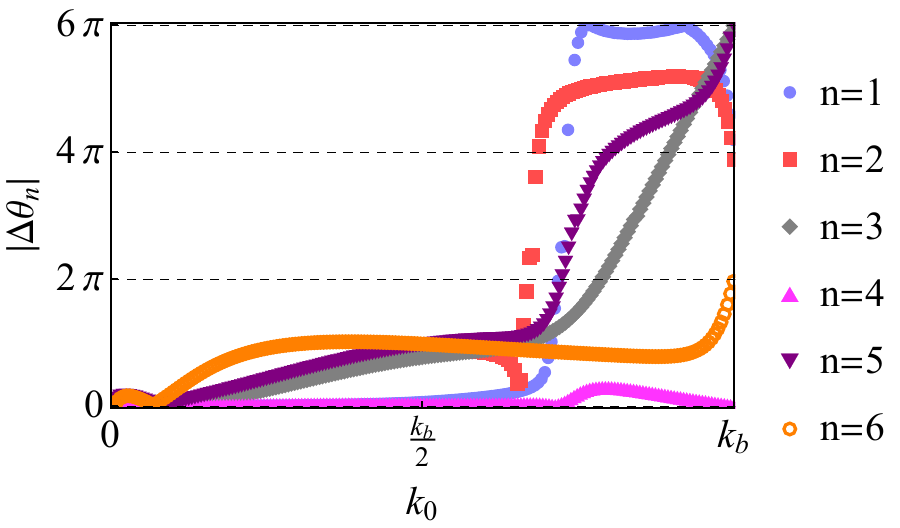}
\label{fig:3c}}
\caption[]{(a) The \emph{ab initio} band structures of Na$_{3}$Bi are plotted along various high- symmetry directions, and the Kramers-degenerate bands are labeled, following an ascending order of energy eigenvalues at the $\Gamma$ point. The linear touching between bands $n=6$ and $n=7$ (purple colored), along the six-fold, screw axis $\Gamma-A$, gives rise to bulk Dirac points (red dot), lying at the Fermi level (dashed line). (b) The Wilson loops are calculated following the hexagonal loop $abcdef$. We increase $k_0$ from zero to $k_b=\pi/a$, when the enclosed area becomes equal to that of hexagonal Brillouin zone (yellow). The gauge invariant eigenvalues of Wilson loops are given by $e^{\pm i \theta_n(k_0)}$, where $n$ is the band index. For topologically non-trivial bands, supporting non-Abelian flux of magnitude $2M\pi$, $|\Delta \theta_n| =| \theta_n(k_0) - \theta_n(0)|$ will interpolate from $0$ to $2M\pi$, as $k_0$ is increased from $0$ to $k_b+\epsilon$, where $\epsilon \geq 0$. (c) For all $xy$ planes, lying between (outside) two Dirac points, the occupied Dirac band, $n=6$, is non-trivial (trivial). (d) At the $k_z=0$ mirror plane, the remote bands $n=1,2(n=3,5)$ also support quantized flux of magnitude $4 \pi$($6 \pi$), demonstrating that the WL method can capture arbitrary flux strength. For more details of higher-winding please consult the supplementary information.}
\label{fig:Pan_3}
\end{figure*}


\begin{figure*}[t]
\centering
\subfigure[]{
\includegraphics[scale=0.18]{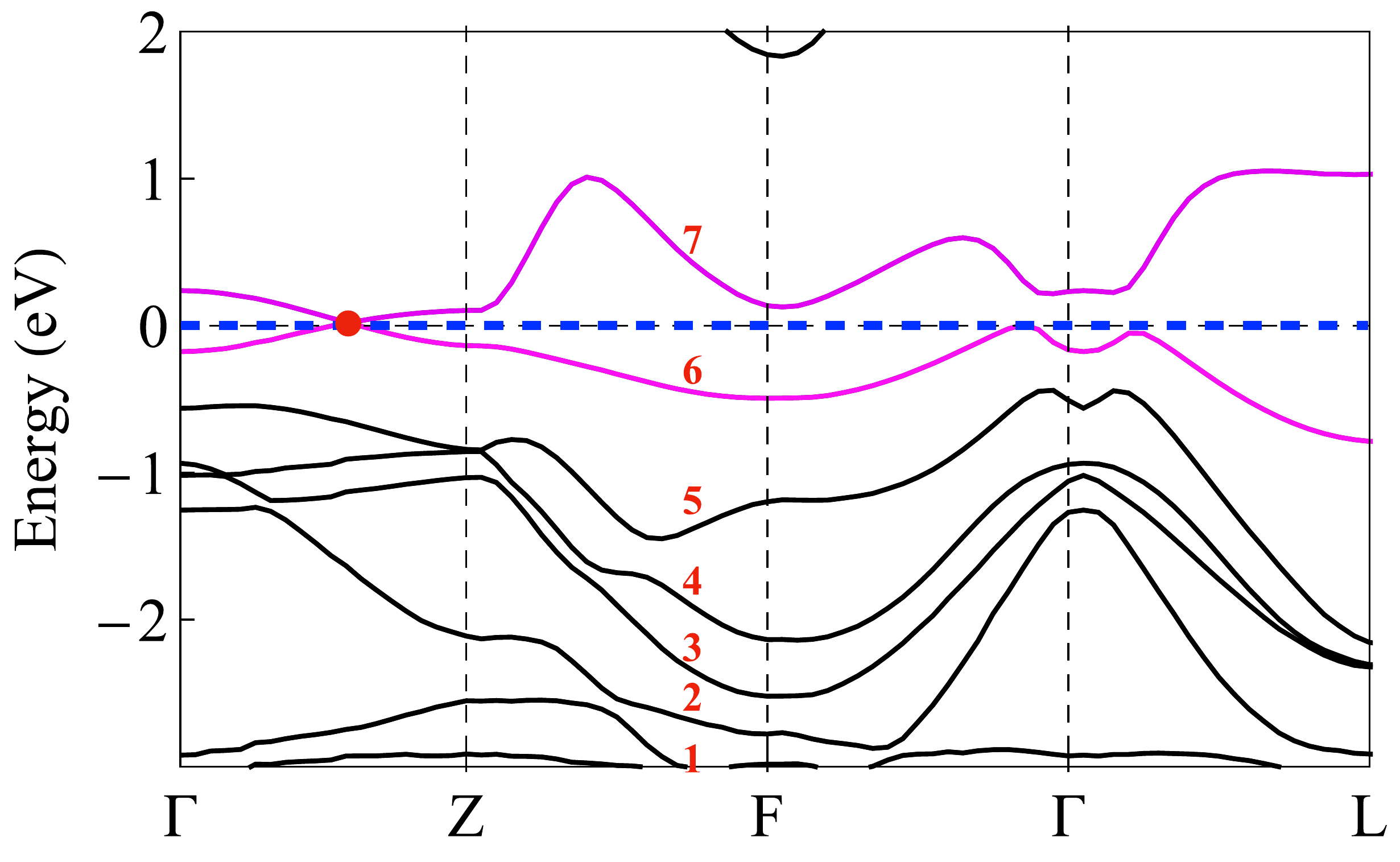}
\label{fig:CuIBands}}
\subfigure[]{
\includegraphics[scale=0.5]{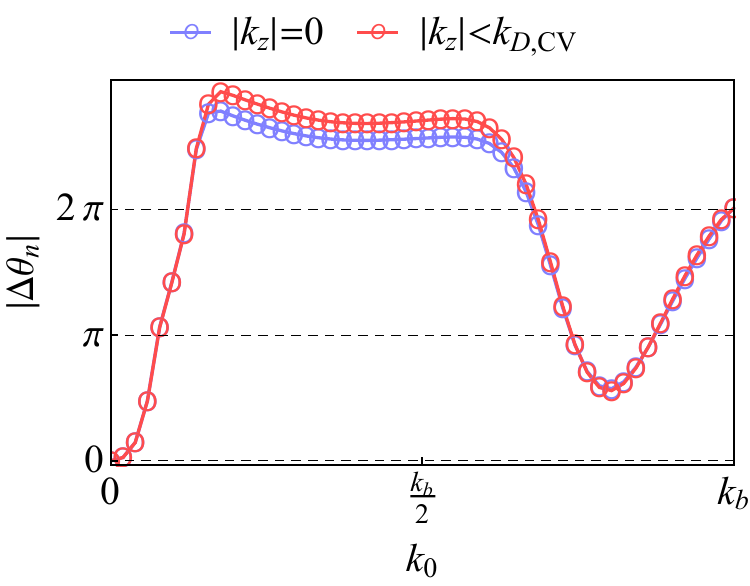}
\label{fig:CuIPWl}}
    \caption{(a) Band structure of $\beta$-CuI in the primitive unit cell along high symmetry path detailed be Le et. al \cite{le2018dirac}. Kramers pairs which touch at the Fermi energy at $\mathbf{k}=(0,0,\pm k_{D})$ to produce Dirac point (bands in Dirac subspace) are colored purple. Dirac point is noted with red dot. (b) Results of Wilson loop calculation at high-symmetry plane, $k_{z}=0$, and generic value of $|k_{z}|<k_{D,CV}$, where $k_{D,CV}$ is the location of the Dirac point in the conventional unit cell for the occupied Dirac band, $n=6$. Wilson loop is calculated as a function of the area enclosed by the Wilson loop path. This calculation indicates that band 6 supports non-Abelian flux of magnitude $2\pi$ for all planes such that $|k_{z}|<k_{D,CV}$.  }
\end{figure*}


\begin{figure*}[t]
\centering
\subfigure[]{
\includegraphics[scale=0.13]{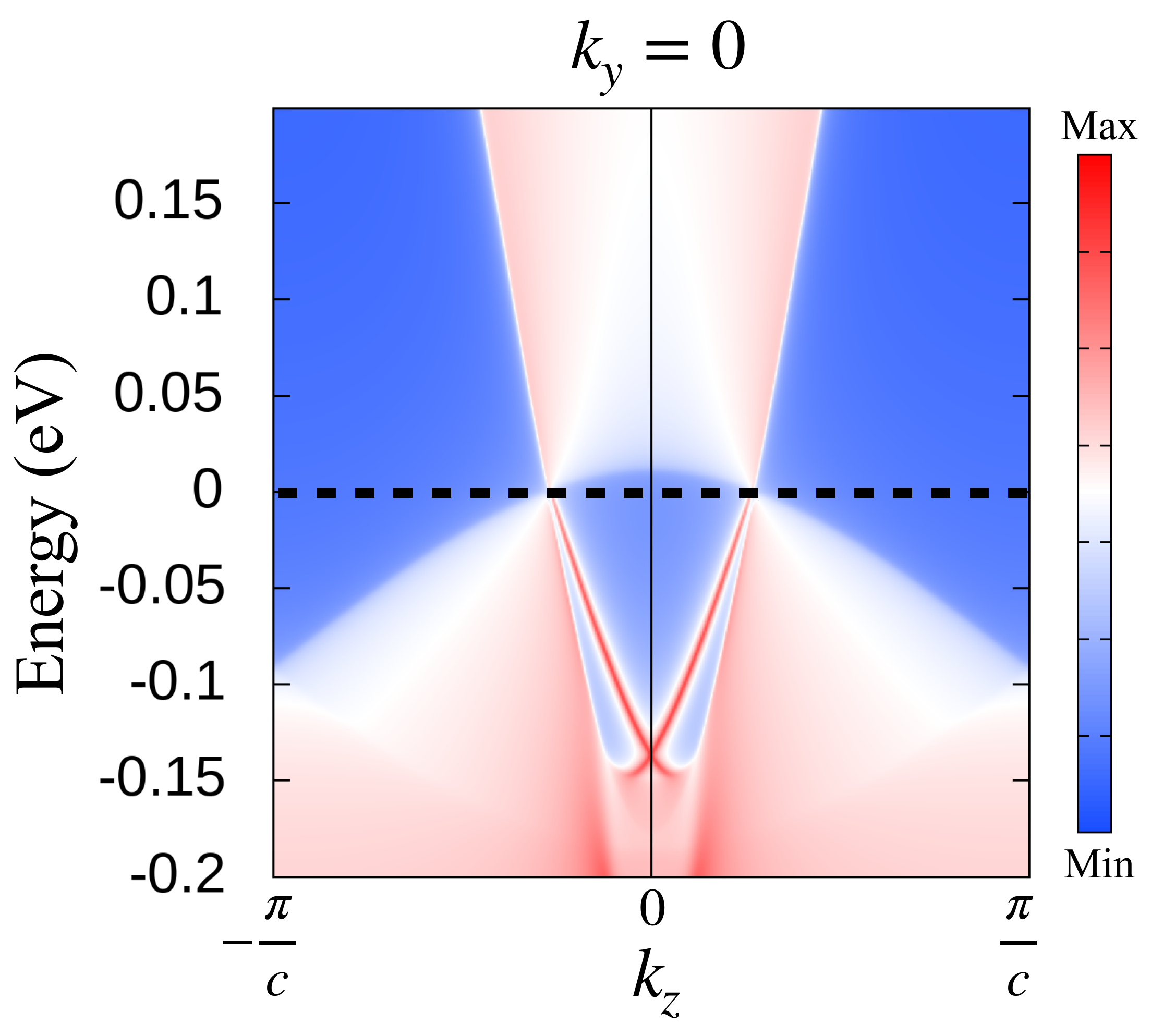}
\label{fig:4a}}
\subfigure[]{
\includegraphics[scale=0.3]{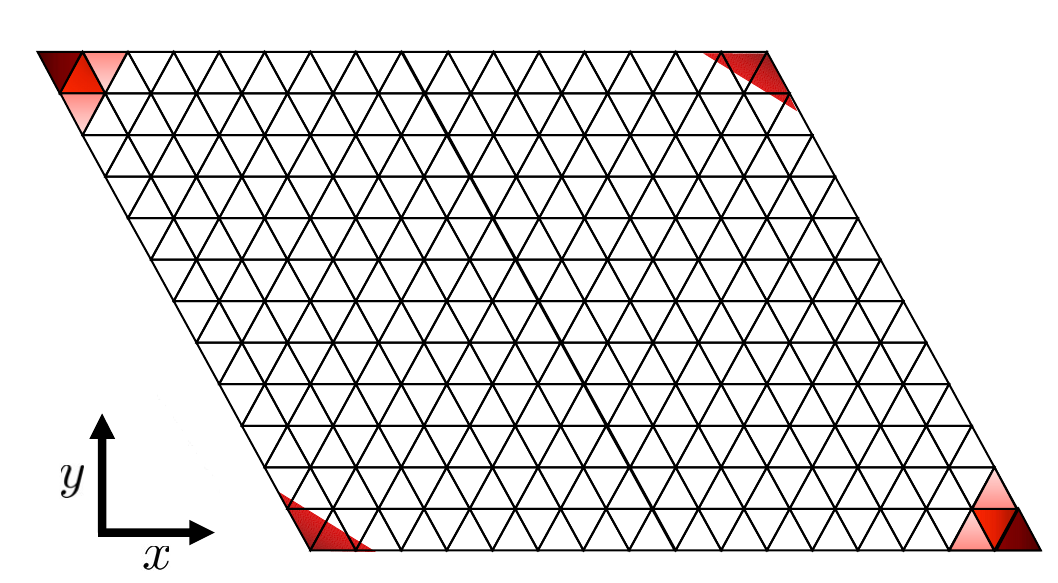}
\label{fig:CorStates}}
\subfigure[]{
\includegraphics[scale=0.13]{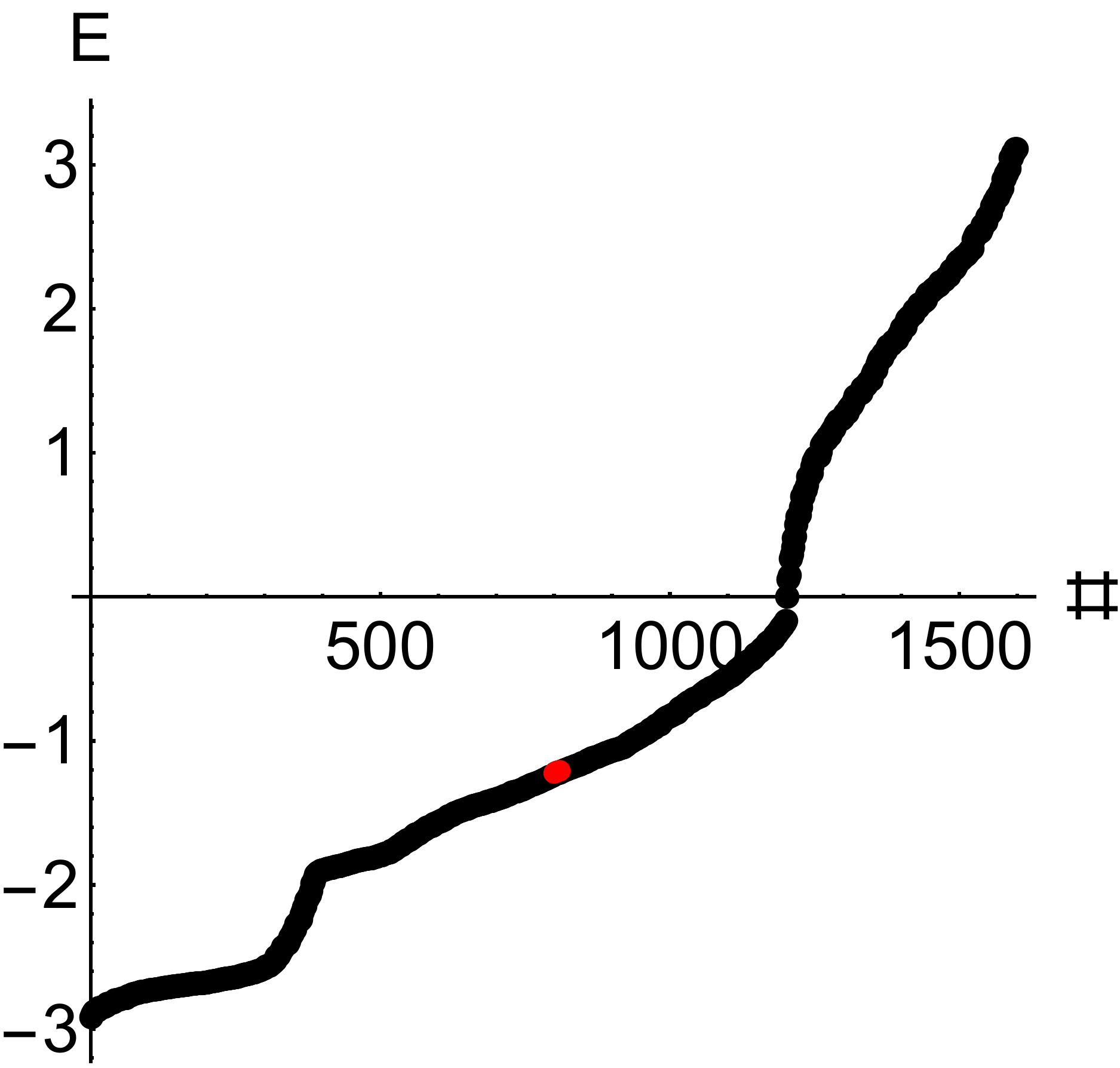}
\label{fig:CornLoc}}
\caption[]{ (a) The (100) edge state dispersion for Na$_{3}$Bi along the $k_z$ axis shows the dependence of gap as a function $k_z$. The normalizability of surface-states breaks down at the images of bulk Dirac points on the surface Brillouin zone. The existence (non-existence) of gapped, surface-states for higher-order topological insulators (trivial insulators) can only be justified by computing quantized, Berry's flux. (b) Schematic of localization pattern for the four corner-localized states at half-filling in eight band tight binding model of Na$_{3}$Bi, setting $|k_{z}|<|k_{D}|$ and solving on a finite slab of $10 \times 10$ primitive unit cells. Darker shading corresponds to stronger localization. The localization pattern confirms that the $xy$ planes between the Dirac nodes, which have been identified as higher order topological insulators, support corner localized states. (c) States resulting from exact diagonalization calculation. Localization shown in (b) corresponds to states colored in red. We note that while these states exist at half-filling, they do not fall at zero energy and are not separated in energy from the bulk states.}
\label{fig:Pan_4}
\end{figure*}


\subsection{Minimal tight binding models}
\par
\begin{figure*}[t]
\centering
\subfigure[]{
\includegraphics[scale=0.35]{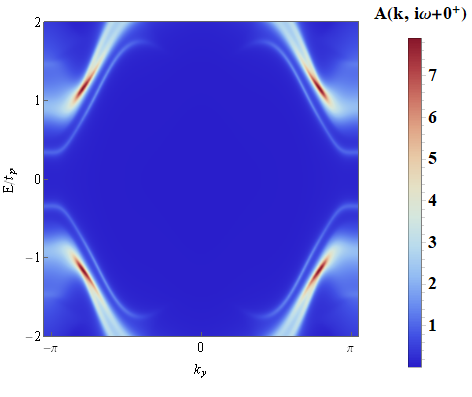}
\label{fig:ED_Cos2k}}
\subfigure[]{
\includegraphics[scale=0.48]{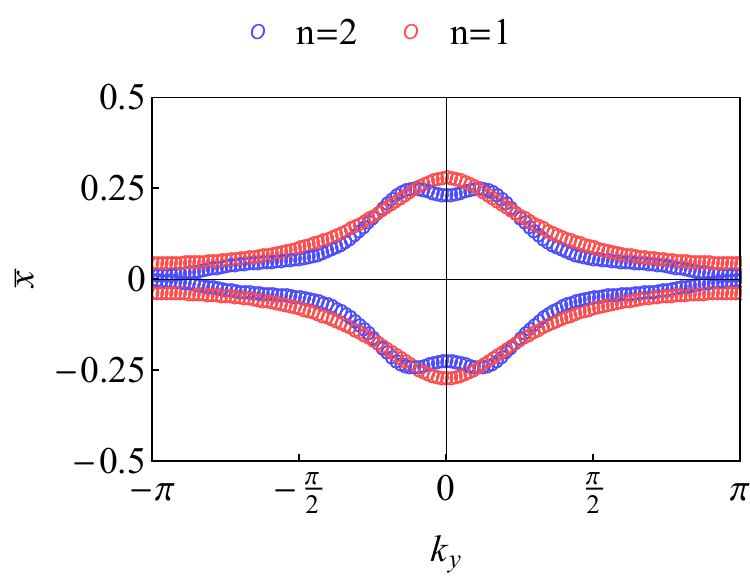}
\label{fig:WCC_Cos2k}}
\subfigure[]{
\includegraphics[scale=0.14]{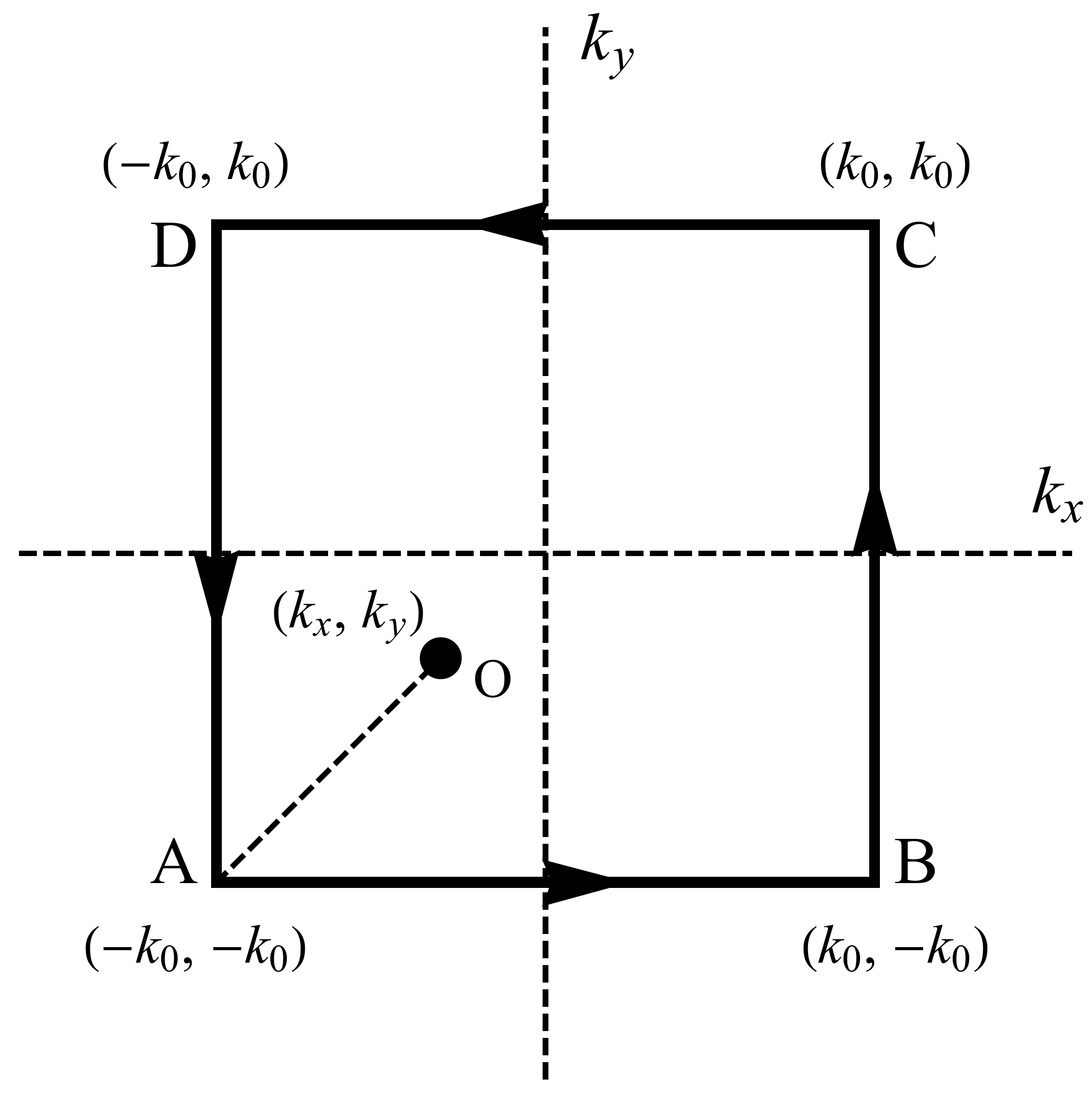}
\label{fig:PWL}}
\subfigure[]{
\includegraphics[scale=0.46]{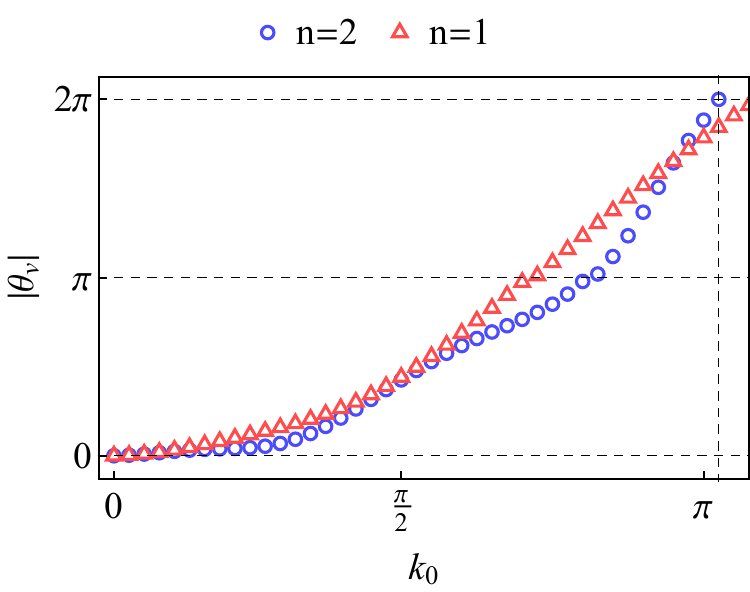}
\label{fig:PWL_Cos2k}}
\caption[]{(a) Spectral density on $(10)$ edge for tight-binding model given by eq. \eqref{eq:TBModel} fixing $n=2$ in arbitrary units. The edge states are gapped when setting $t_{p,1}=t_{p,2}=t_{s}=t_{d}=1$, and considering a semi-infinite slab along the $x$ direction with periodic boundary conditions along $k_{y}$. (b) Wannier center charges (WCCs), $\bar{x}(k_{y})$, fixing $n=1,2$ for identical choice of hopping parameters. The WCC spectra does not display winding in either case, indicating trivial topology. (c) The $\mc C_4$-symmetric path $ABCD$ for computing Wilson loops.  (d) Results of WL following a $C_{4}$ symmetric closed contour. As the area enclosed by the contour approaches the area of the Brillouin zone ($k_{0}=\pi$), the flux converges to $|2\pi|$ for $n=2$, while a contour slightly larger than the area of the zone must be used when $n=1$, capturing the non-trivial topology. }
\label{fig:Pan_1}
\end{figure*}

 
To demonstrate the necessity of the proposed in-plane Wilson loop method, consider the tight-binding model, $H(\mathbf{k})_{n}=\sum_{j=1}^{5}N^{n}_{j}(\mathbf{k})\Gamma_{j}$, where, 
\begin{eqnarray}\label{eq:TBModel}
    \mathbf{N}(\mathbf{k})^{n}&=&\{t_{p}\sin k_{x},t_{p}\sin k_{y},t_{d}\sin k_{x}\sin k_{y},\lambda_{n}(\mathbf{k}), \nn \\ && t_{s}(3/2- \cos k_{x}-\cos k_{y})\}.
\end{eqnarray}
The lattice constants have been set to unity, $t_{j}$'s are hopping parameters with units of energy, and $\Gamma_j$'s are five, mutually anti-commuting, $4 \times 4$ matrices, such that $\{\Gamma_i, \Gamma_j\}=2\delta_{ij}$. Specifically we set, $\Gamma_{i=1,2,3}=\tau_{i}\otimes \sigma_{1},\; \Gamma_{4}=\tau_{2}\otimes \sigma_{0}, \Gamma_{5}=\tau_{3} \otimes \sigma_{0}$, where $\sigma_{i=1,2,3}(\tau_{i=1,2,3})$ $\tau_{0}$ and $\sigma_{0}$ are the Pauli matrices and two $2 \times 2$ identity matrices respectively. We then define $\lambda_{n}(\mathbf{k})=t_{d}(\cos (nk_{x})-\cos (nk_{y}))$. Parity symmetry is given by $\mathcal{P}=\tau_{3} \otimes \sigma_{3}$, such that $\mathcal{P}^{\dagger}H(\mathbf{k})\mathcal{P}=H(-\mathbf{k})$. Despite the presence of parity symmetry, assignment of a Fu Kane strong TI index is inconclusive. This is clearly seen in the case $n=2$; at the TRIM locations only $N_{5}(\mathbf{k})\neq0$, thus $\mathcal{P}$ and $H(\mathbf{k})_{n}$ are both diagonal. The constituent states of each Kramers pair are thus labeled by opposite eigenvalues of $\mathcal{P}$ at each TRIM location. If we were to rely on the WCC spectra for diagnosis of bulk topology, eigenvalues of $\text{Im(Ln(}W_{x}(k_{y})))/(2\pi)$ denoted $\bar{x}(k_{y})$, we would conclude that the model is trivial, see Fig. \eqref{fig:WCC_Cos2k}. It is only by computing the WL following the path ABCD shown in Fig. \eqref{fig:PWL}, with results shown in Fig. \eqref{fig:PWL_Cos2k}, that it is clear this model supports Berry's flux, for which quantization to $|2\pi|$ is commensurate with the BZ when $n=2$. If $n=1$, multiple components of $\mathbf{N}(\mathbf{k})$ survive at the TRIM locations obscuring winding of the five component vector field. This is the reason a WL along the boundary of the BZ is not quantized. The non-Abelian phase is exhibiting a twisted boundary condition. However, by expanding the area enclosed by the WL contour, a full winding from $0\rightarrow 2\pi$ can again be identified. Furthermore, a corresponding quantized flux in the BZ can be identified for arbitrary $n$ via Abelian gauge fixing with respect to the rotation generator $\Gamma_{12}$, as detailed in Tyner et. al \cite{tyner2020topology}. Nevertheless, WL has the advantage of being a basis-agnostic method which can be readily applied to \emph{ab initio} data using existing software. 

\subsection{Higher winding numbers}
\begin{figure*}[t]
\centering
\subfigure[]{
\includegraphics[scale=0.5]{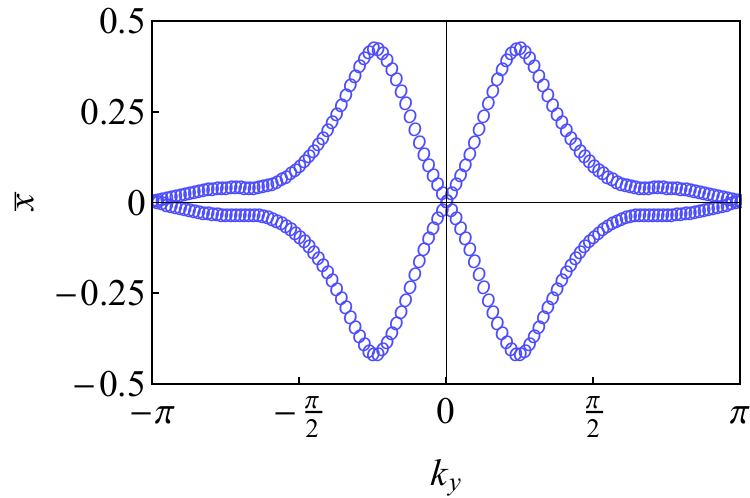}
\label{fig:WCCDouble}}
\subfigure[]{
\includegraphics[scale=0.35]{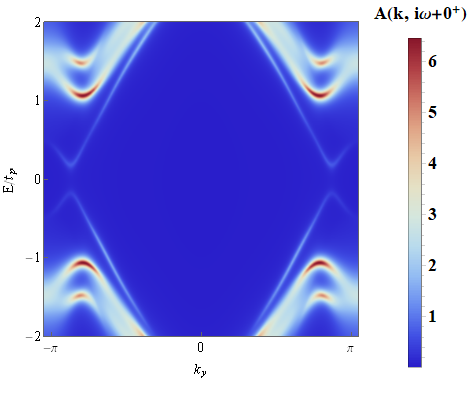}
\label{fig:StatesGF}}
\subfigure[]{
\includegraphics[scale=0.15]{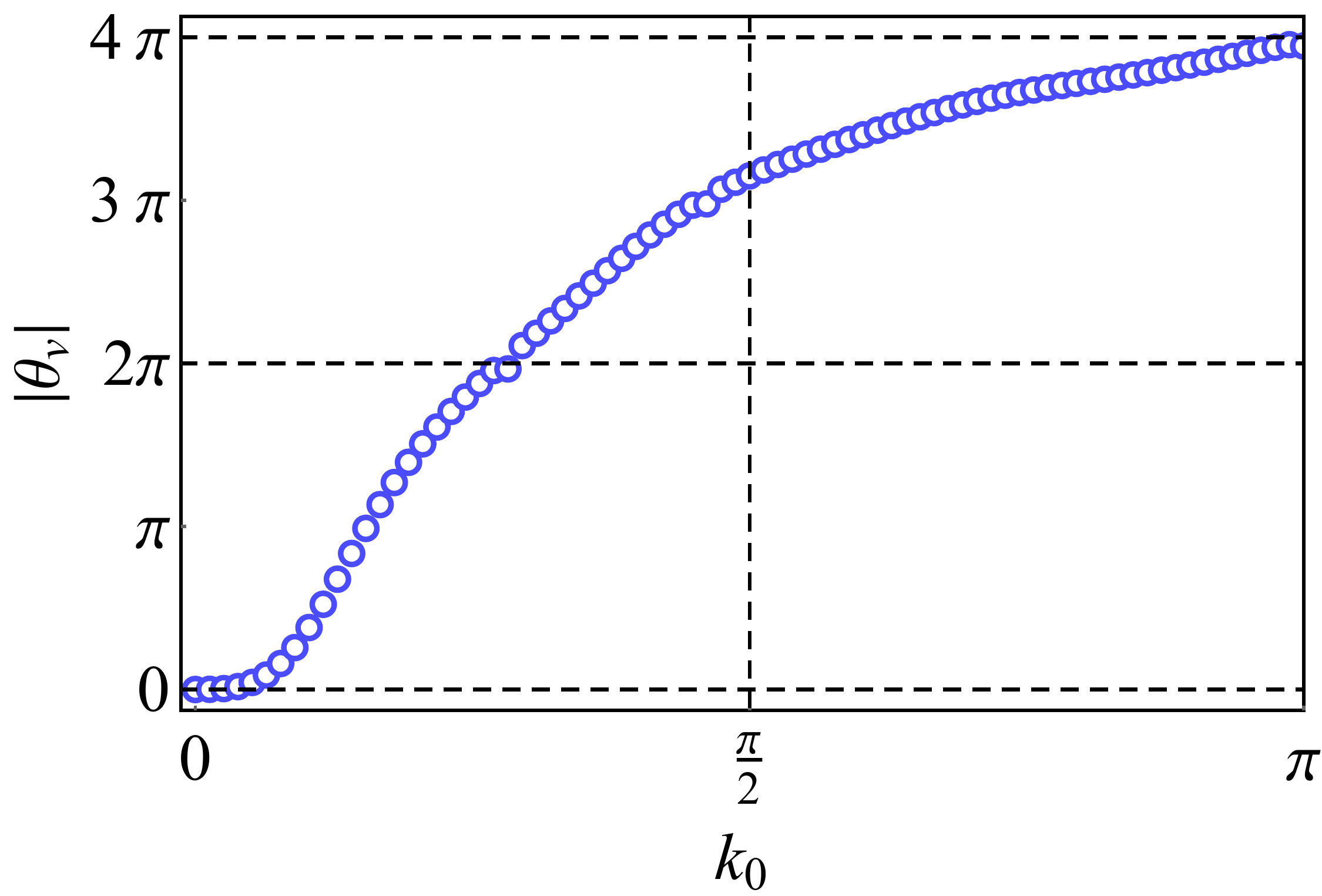}
\label{fig:PWLDouble}}
\subfigure[]{
\includegraphics[scale=0.4]{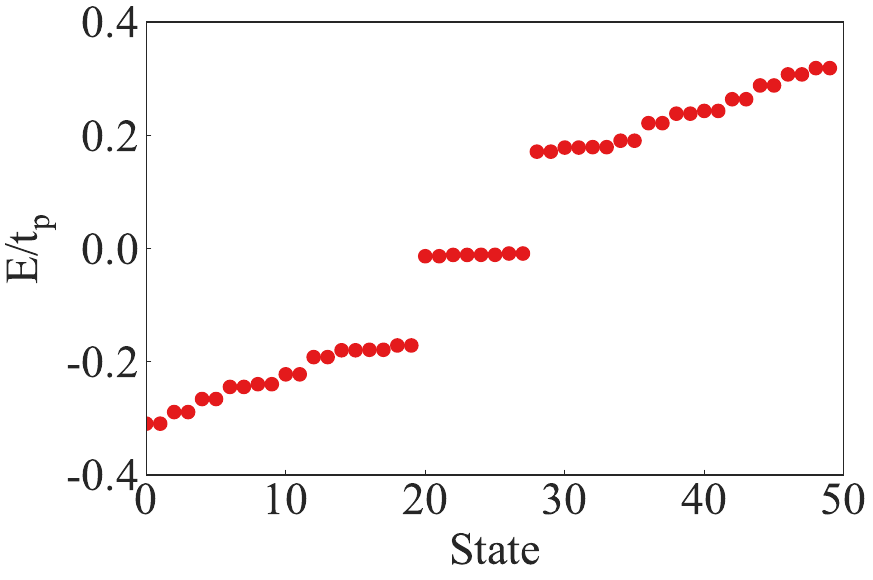}
\label{fig:StatesDouble}}
\subfigure[]{
\includegraphics[scale=0.4]{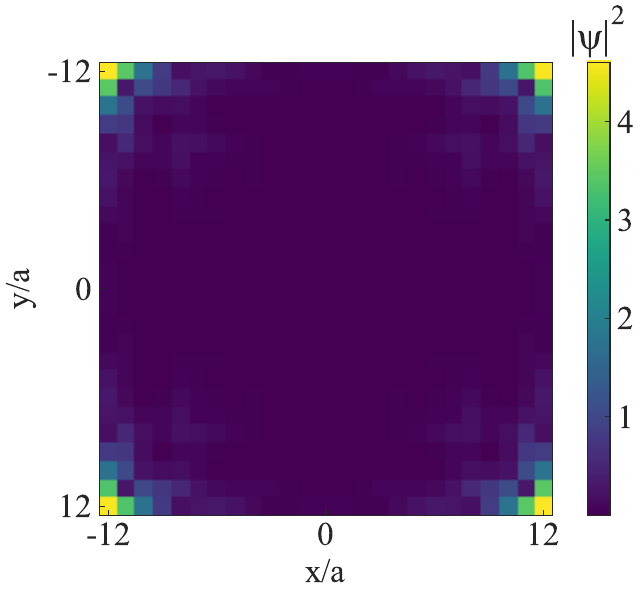}
\label{fig:StatesLoc}}
\caption[]{(a) Wannier center charge (WCC) spectra for the tight-binding model given by eq. \eqref{eq:doublewind}, fixing all hopping parameters to be equal. The WCCs indicate trivial first-order topology and no gapless edge states. (b) Spectral density on (10) surface of eq. \eqref{eq:doublewind}, considering a semi-infinite slab geometry. The results indicate the lack of gapless edge states in accordance with the WCC spectra. (c) Results of WL calculation for eq. \eqref{eq:doublewind} as a function of the area enclosed by the contour. The WL indicates a quantized non-Abelian flux of $|4\pi|$. (d) When considering a finite slab of 25 unit cells along the x and y directions, the fifty smallest magnitude states are shown. There exists eight mid-gap states at zero energy. (e) Localization of the eight mid-gap states shown in (d), indicating that these are corner localized.}
\label{fig:Double}
\end{figure*}

It is possible to construct a tight-binding model exhibiting quantized non-Abelian flux greater than $2\pi$. As an example, we consider the Bloch Hamiltonian, $H(\mathbf{k})=\sum_{j=1}^{5}N_{j}(\mathbf{k})\Gamma_{j}$,  where, 
\begin{eqnarray}\label{eq:doublewind}
    \mathbf{N}(\mathbf{k})&=&\{\lambda^{1}_{1}(\mathbf{k}),\lambda^{1}_{2}(\mathbf{k}),\lambda^{2}_{1}(\mathbf{k}),\lambda^{2}_{2}(\mathbf{k}), \nn \\ && t_{s}(3/2-\cos k_{x}- \cos k_{y})\},
\end{eqnarray}
defining $\lambda^{1}_{n}=t_{d,n}\sin (n k_{x})\sin (n k_{y})$ and $\lambda^{2}_{n}=t_{d,n}(\cos (n k_{x})-\cos (n k_{y}))$. We will work with $t_{d,1}=t_{d,2}=t_{s}$. For this model, parity symmetry is generated by the identity matrix, precluding assignment of a Fu-Kane $Z_{2}$ index. Calculating WCCs, $\bar{x}(k_{y})$, we find the result shown in Fig. \eqref{fig:WCCDouble}, indicating a trivial configuration. This result is in-line with the absence of gapless edge states as shown in Fig. \eqref{fig:StatesGF}. When applying open boundary conditions in two-dimensions, we note the existence of mid-gap corner states, shown in Fig. \eqref{fig:StatesDouble} and \eqref{fig:StatesLoc}. However, the nested Wilson loop calculations yields trivial results in this model. As such, we are left without a method capable of diagnosing topology under periodic boundary conditions. 
\par 
Implementing the WL, we find the results shown in Fig. \eqref{fig:PWLDouble}. These results demonstrate beautiful quantization of non-Abelian flux within the Brillouin zone to a magnitude of $|4\pi|$. 

Contrast to nested Wilson loops:
The method of nested Wilson loops (NWLs)\cite{benalcazar2017,schindler2018} has been developed to identify higher-order topological insulators. More precisely, the method of NWLs is used to identify bulk-boundary correspondence for hinge/corner localized states when edge states are gapped. Application of this method to the tight-biding model given by eq. \eqref{eq:TBModel} in the main-body, yields a non-trivial result when all hopping parameters are non-zero. As this model supports mid-gap corner states, the NWL can be used to describe the bulk boundary correspondence of these states. NWLs will fail when $t_{d}=0$ as gapless edge states are present. WL does not suffer from this dependence on edge/corner states and can be applied successfully for arbitrary hopping parameters. More importantly, in the model given by eq. \eqref{eq:doublewind}, both WCCs and NWL fail to indicate non-trivial topology despite the presence of mid-gap corner states. This leaves the question of bulk-boundary correspondence an open one under the current paradigm of topological classification. Only WL is capable of yielding information of non-trivial topology under periodic boundary conditions, emphasizing non-Abelian flux as the universal signature of non-trivial topology under periodic boundary conditions. 

\section{$C_{n=3,6}$ symmetric tight-binding models}
In order to demonstrate how implementation of WL can be extended from the four-fold rotationally symmetric system, to a three- and six-fold system, in a controlled manner, we introduce two tight binding Hamiltonians of two Kramers degenerate bands. The first is a two-dimensional topological insulator with six-fold rotational symmetry, described by the Hamiltonian, $H=\sum_{k}\mathbf{\Psi}^{\dagger}(\mathbf{k})\hat{H}(\mathbf{k})\mathbf{\Psi}(\mathbf{k})$, where $\mathbf{\Psi}(\mathbf{k})$ is a four-component spinor. The Bloch Hamiltonian operator is written as $\hat{H}(\mathbf{k})=\sum_{j=1}^{5}N_{j}(\mathbf{k})\Gamma_{j}$, where the five component vector $\mathbf{N}(\mathbf{k})$ contains details of the band structure and the $\Gamma$ matrices have been previously defined. We will consider, 
\begin{eqnarray}\label{eq:c6_Mod}
   && \mathbf{N}(\mathbf{k})=t\{(2\sin(k_{1}+k_{2})+\sin k_{1}+\sin k_{2})/3, (\sin k_{1} \nn \\ && -\sin k_{2})/\sqrt{3},  4(\cos k_{1}+\cos k_{2}  -2\cos (k_{1}+k_{2}))/3, \nn \\ && 4(\cos k_{2}-\cos k_{1})/\sqrt{3}, \Delta- (\cos(k_{1}+k_{2})+\cos k_{1}  \nn \\ && + \cos k_{2})\},
\end{eqnarray}
where t is a hopping parameter with units of energy that will be fixed such that $t=1$ in all calculations, and $k_{j}=\mathbf{k}\cdot \mathbf{a}_{j}$, for lattice vectors, $\mathbf{a}_{1}=a(\sqrt{3},1)/2$, $\mathbf{a}_{2}=a(-\sqrt{3},1)/2$. We have set $a=1$ such that there exists unit lattice spacing in all directions. 
\par
\begin{figure*}[t]
\centering
\subfigure[]{
\includegraphics[scale=0.5]{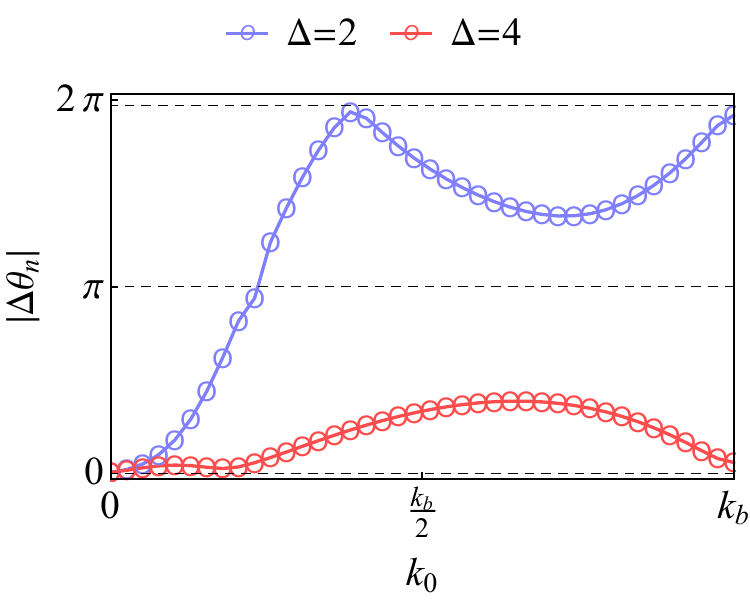}
\label{fig:C6}}
\subfigure[]{
\includegraphics[scale=0.5]{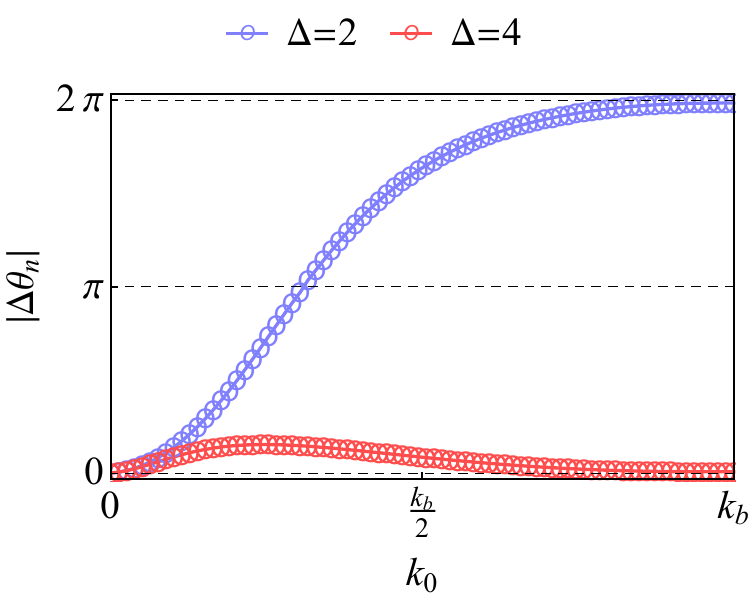}
\label{fig:C3}}
\caption[]{ (a) Results of WL calculation for $C_{6}$ symmetric model given by eq. \eqref{eq:c6_Mod} following the path shown in Fig. \eqref{fig:3a} for distinct values of the non-thermal band parameter $\Delta$. Results demonstrate that this model supports non-Abelian flux of magnitude $2\pi$ for $\Delta=2$, and is trivial when $\Delta=4$. (c) Results of WL calculation for the $C_{3}$ symmetric model given by eq. \eqref{eq:c3_mod} for distinct values of $\Delta$. As in the $C_{6}$ case, the results demonstrate that this model supports non-Abelian flux of magnitude $2\pi$ for $\Delta=2$, and is trivial when $\Delta=4$. }
\label{fig:C3C6}
\end{figure*}

 
Upon fixing $\Delta=2$, we compute the WL, following the closed, directed path shown in Fig. \eqref{fig:3a}, with the results displayed in Fig. \eqref{fig:C6}. This procedure is then repeated fixing $\Delta=4$. indicating that for $\Delta=2(4)$ the model is in a topological (trivial) phase supporting quantized flux of magnitude $|2\pi|(0)$.
\par 
Next, we turn to a three-fold symmetric example, altering $ \mathbf{N}(\mathbf{k})$
in eq. \eqref{eq:c6_Mod}, to the form, 
\begin{eqnarray}\label{eq:c3_mod}
   && \mathbf{N}(\mathbf{k})= t\{\sin k_{1}, -\sin k_{2}, -\sin (k_{1}-k_{2}),0, \nn \\ && (\Delta+\cos k_{1}+\cos k_{2}+ \cos (k_{1}+k_{2}))\},
\end{eqnarray}
where t is again a hopping parameter with units of energy, fixed to unity, and $k_{j}=\mathbf{k}\cdot \mathbf{a}_{j}$, for lattice vectors, $\mathbf{a}_{1}=a(\sqrt{3},1)/\sqrt{6}$, $\mathbf{a}_{2}=a(\sqrt{3},-1)/\sqrt{6}$. The three-fold environment is particularly important to examine as three-fold planes do not support mirror symmetry when embedded in a three-dimensional structure. Thus, although parity symmetry is given by $\mathcal{P}=\Gamma_{5}$, and $Z_{2}$ invariant $\nu_{0}=+1$ can be assigned, such systems have escaped classification via quantized flux. WLs are calculated following the same procedure defined previously, with the results shown in Fig. \eqref{fig:C3}. These results demonstrate that when $\Delta=2(4)$ the model is in a topological (trivial) phase, supporting flux $|2\pi|(0)$.

\section{Details of Na$_{3}$Bi analysis}
\par 
All first-principles calculations based on the density-functional theory are performed using the Vienna \textit{ab initio} simulation package~\cite{Kresse1996VASP,kresse1999VASP}, and the exchange-correlation potentials use the Perdew-Burke-Ernzerhof (PBE) parametrization of the generalized gradient approximation~\cite{Perdew1996}. An 11$\times$11$\times$7 grid of $\bs{k}$ points and a plane-wave cutoff energy $520$ eV are used for self-consistent field calculations. All calculations incorporate the effects of spin-orbit coupling. The qualitative features of DSM phase of Na$_{3}$Bi have been well characterized with the first principles calculations of band structures and various spectroscopic, and transport measurements\cite{wang2012dirac,liu2014discovery,Xu294,kushwaha2015bulk,xiong2015evidence,liang2016electronic}.
\par
\begin{figure*}[t]
\centering
\subfigure[]{
\includegraphics[scale=0.26]{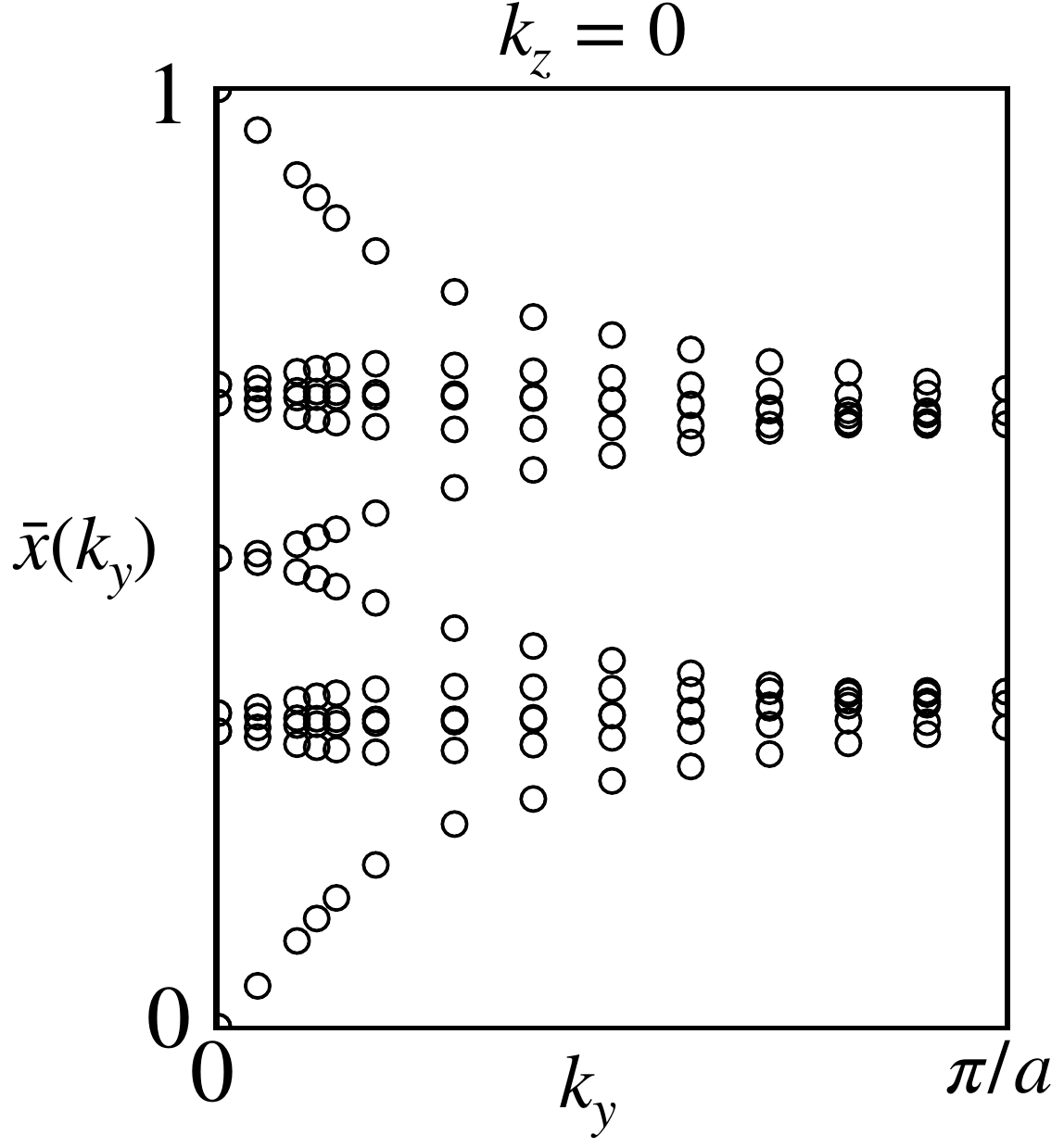}
\label{fig:7a}}
\subfigure[]{
\includegraphics[scale=0.26]{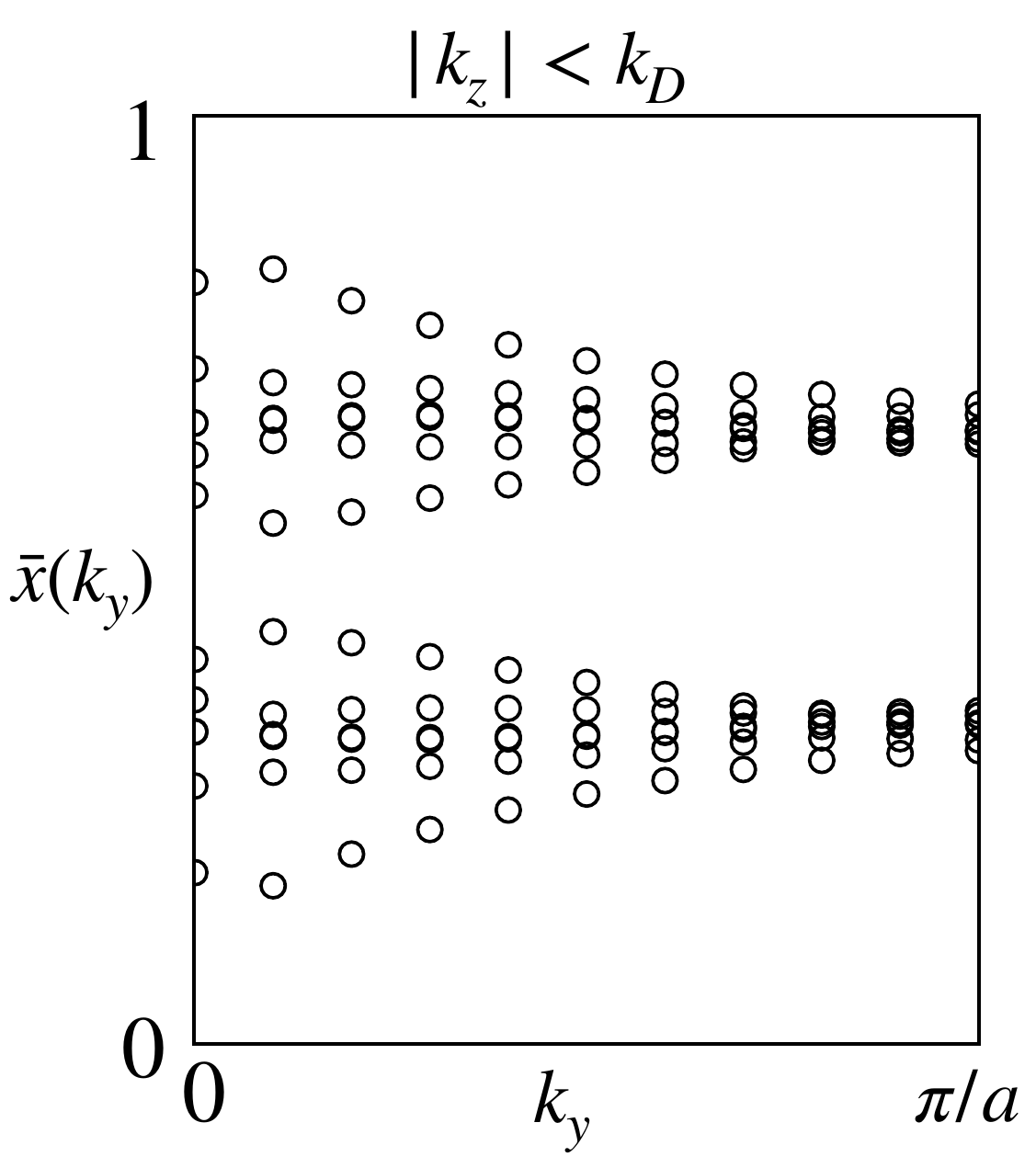}
\label{fig:7b}}
\subfigure[]{
\includegraphics[scale=0.26]{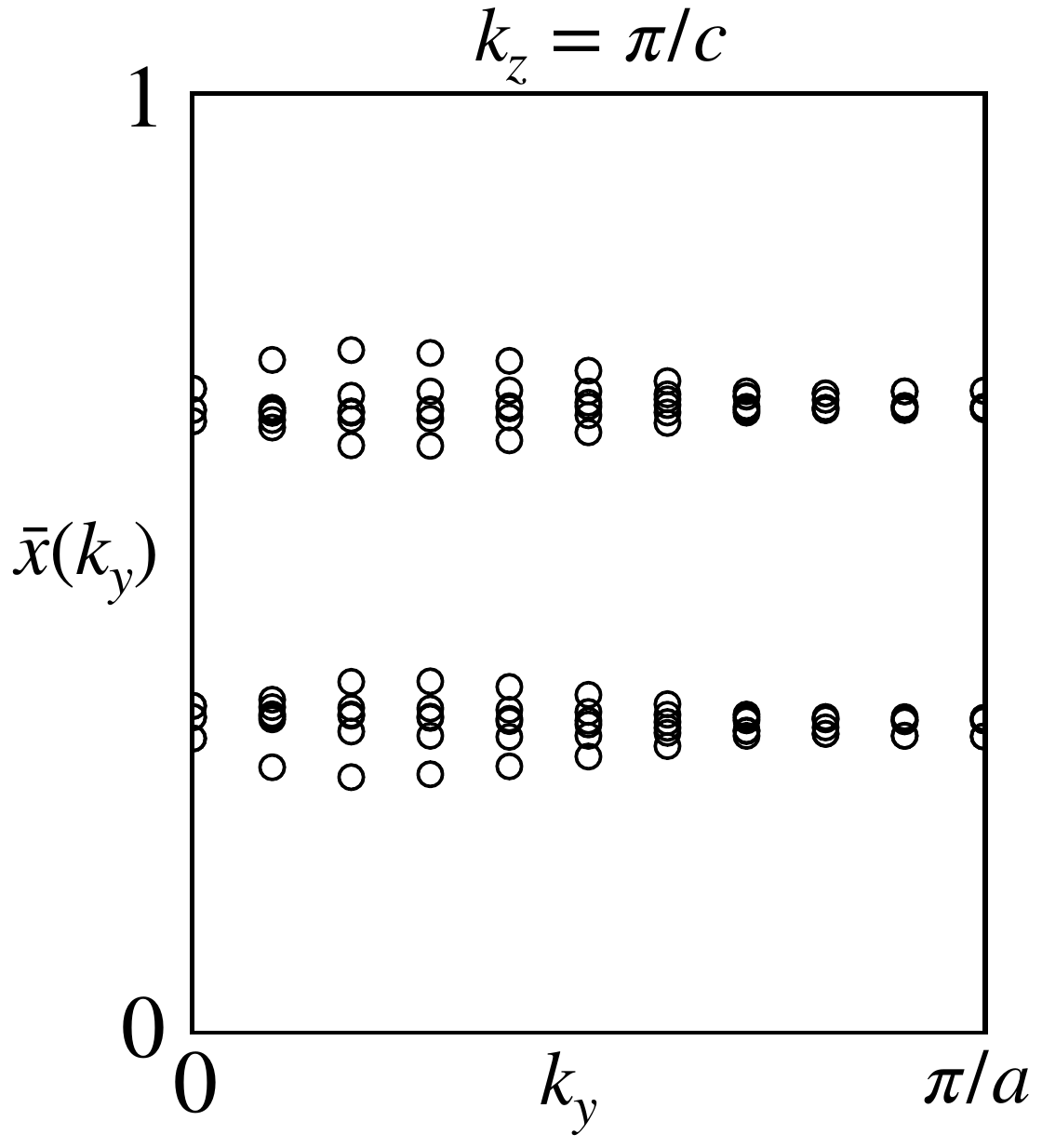}
\label{fig:7c}}
\caption[]{The hybrid Wannier centers or spectra of Polyakov loops of non-Abelian Berry's connections of occupied bands $n=1$ through $n=6$ in Na$_{3}$Bi for (a) the $k_z=0$ mirror plane, (b) the higher-order topological insulators with $|k_z| < k_D$, and (c) the topologically trivial mirror plane $k_z=\pi/c$. The gapless Wannier spectra and partner-switching allow us to identify the $k_z=0$ plane as a first-order topological insulator with a $Z_2$ invariant. This is consistent with the three, occupied bands $n=3$, $5$, and $6$ supporting non-trivial, first homotopy classification, and the existence of zero-energy, helical edge states for this plane. The higher-order topological insulators and the trivial insulators exhibit gapped Wannier spectra, as none of the occupied bands support non-trivial, first homotopy classification. Therefore, the hybrid Wannier centers cannot distinguish between the trivial and the higher-order topological insulators.}
\label{fig:WCCOCC}
\end{figure*}

\begin{figure*}[t]
\centering
\subfigure[]{
\includegraphics[scale=0.14]{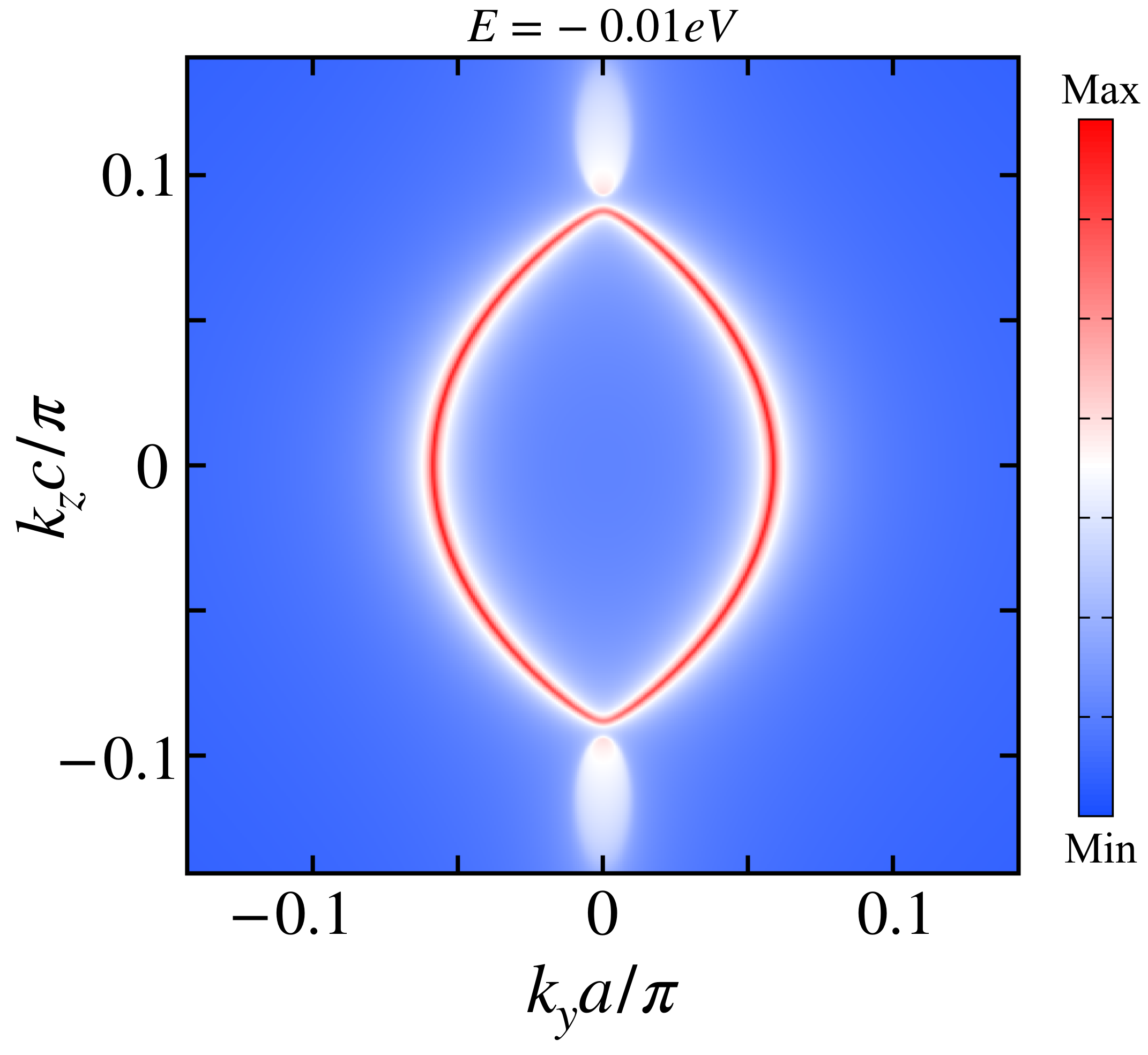}
\label{fig:8a}}
\subfigure[]{
\includegraphics[scale=0.14]{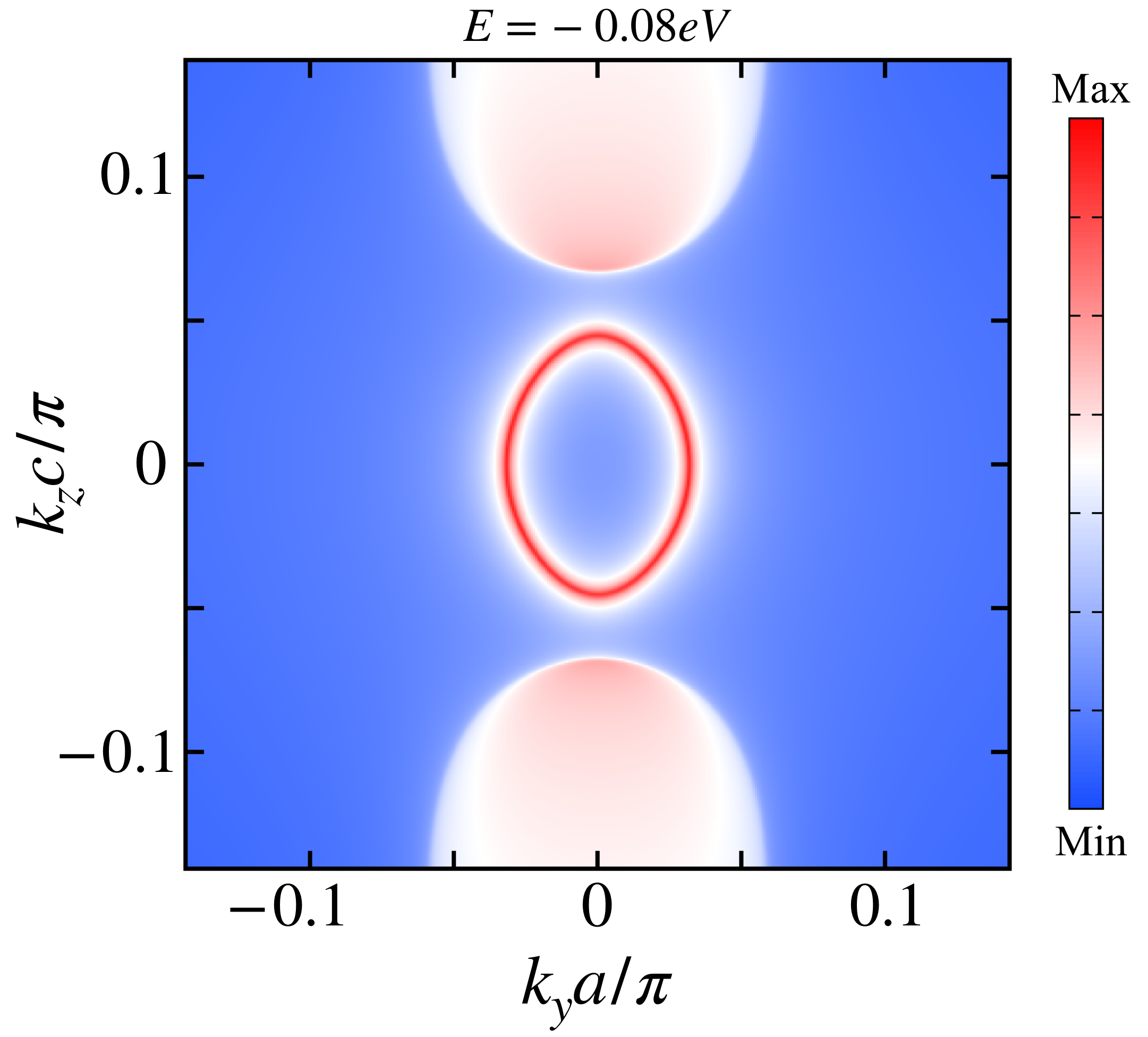}
\label{fig:8b}}
\subfigure[]{
\includegraphics[scale=0.14]{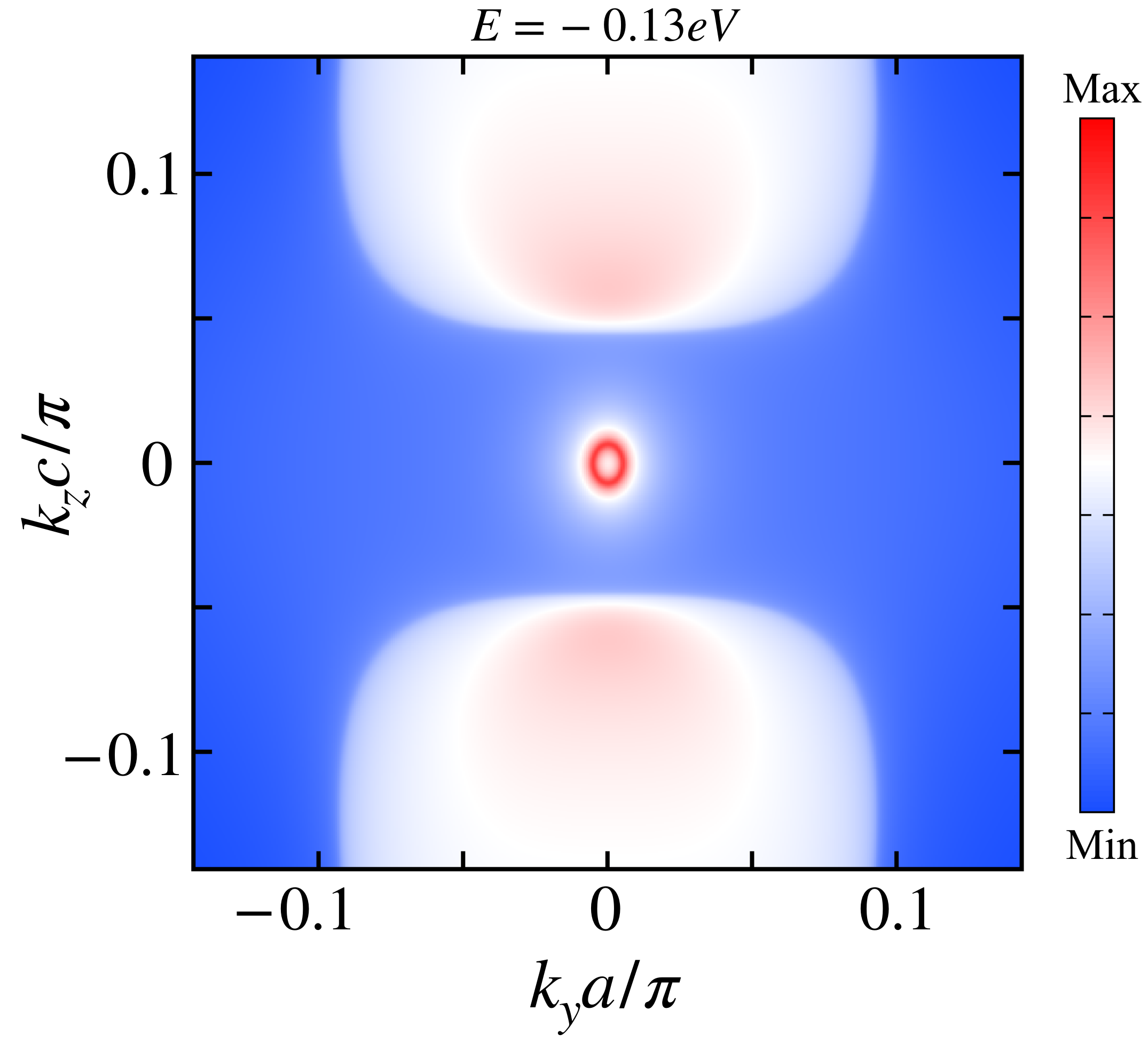}
\label{fig:8c}}
\caption[]{The spectral intensity of surface states of Na$_{3}$Bi for different choice of reference energies. The constant energy cuts correspond to the Fermi surface of surface-bands, instead of topologically protected, loci of zero-energy states or helical Fermi arcs. Unlike the constant $k_z$ slices, the fermiology of surface-states cannot reveal clear signatures of bulk topology. }
\label{fig:GFSlice}
\end{figure*}

 
When performing a topological analysis of Dirac and Weyl semimetals, it is common to calculate the WCCs of all occupied states, we present such an analysis in Fig. \eqref{fig:WCCOCC}. At the high-symmetry planes the behavior of WCCs can be used to determine the strong TI invariant as well as the mirror Chern number\cite{KhalafSymm,Kruthoff2017,bradlyn2017topological,po2017symmetry,CanoBuildingBlocks2018,vergniory2019complete,zhang2019catalogue,tang2019efficient,tang2019comprehensive, vergniory2021all,xu2020high,elcoro2020magnetic,Bouhon2021,Lange2021}, however, as shown in the main body, there are instances where this correspondence breaks down. Further, only by performing a WL can one obtain topological information at a generic plane. It is also common to search for helical Fermi arcs as a signature of the bulk topology, examining the surface spectral density at a fixed energy as a function of momenta, as seen in Fig. \eqref{fig:GFSlice}. Our results demonstrate that such depictions of the surface modes can lead to the misconception that genuine helical Fermi arcs are present in a Dirac semimetal. We thus emphasize the importance of bulk topological classification using WLs. This method can be used for establishing the topological universality class of DSMs in various compounds such as Cd$_3$As$_2$~\cite{wang2013three}, BiAuBi-family~\cite{PhysRevB.91.205128}, Cu$_3$PdN~\cite{PhysRevLett.115.036807}, LiGaGe-family~\cite{Du2015},PdTe$_2$ ~\cite{Huang2016}, $\beta'$-PtO$_2$~\cite{kim2019,wieder2020strong}, VAl$_3$~\cite{chang2017type}, $\beta$-CuI~\cite{le2018dirac}, KMgBi~\cite{wieder2020strong,le2017three}, FeSn~\cite{lin2020dirac}.

\par
\section{Details of $\beta$-CuI analysis}
$\beta$-CuI belongs to space group $R\bar{3}m$ and was recently proposed as a type-I Dirac semimetal \cite{le2018dirac}. The bulk DPs fall along the $k_{z}$ axis, and are thus protected by the three-fold rotational symmetry of the $k_{z}$ axis as well as $\mathcal{PT}$ symmetry. First-principles calculations are carried out using the same packages implemented for Na$_3$Bi. The BZ is sampled with an $8\times8\times8$ grid of $\mathbf{k}$ points and a plane-wave cutoff of 500 eV is used for self-consistent field calculations. The calculated band structures within the energy window $-3eV$ and $2eV$ is displayed in Fig. \eqref{fig:CuIBands} with the Kramers-degenerate bands labeled according to their energy at the $\Gamma$ point, with $E_{n}(0)<E_{n+1}(0)$.
\emph{Bulk-edge correspondence:}
In Le et. al \cite{le2018dirac}, it was shown that $\beta$-CuI does not support Fermi arcs, but rather gapped edge states for all generic planes satisfying $|k_{z}|<k_{D}$ when open-boundary conditions are applied perpendicular to the direction of nodal separation. We can thus conclude that the bulk-edge correspondence of these states is given by the presence of quantized non-Abelian flux.

\newpage

\begin{figure*}[t]
\centering
\subfigure[]{
\includegraphics[scale=0.4]{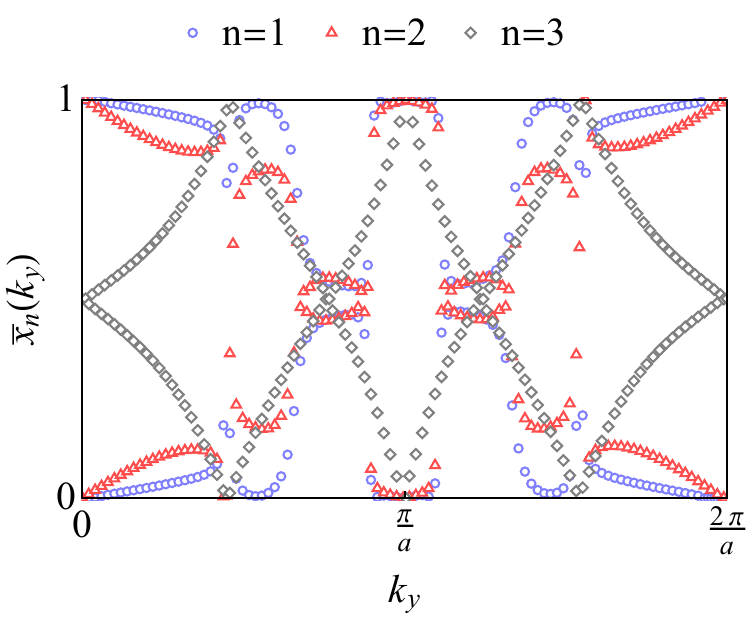}
\label{fig:7a}}
\subfigure[]{
\includegraphics[scale=0.4]{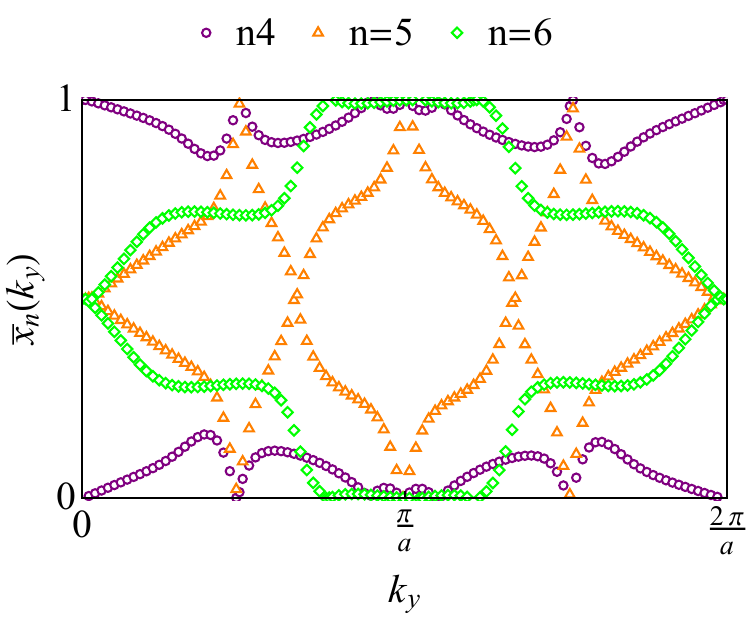}
\label{fig:7b}}
\subfigure[]{
\includegraphics[scale=0.4]{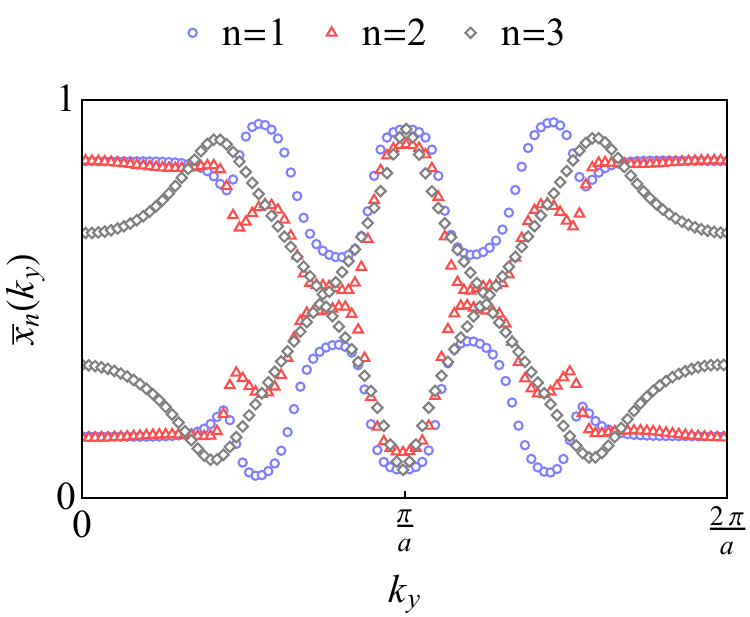}
\label{fig:7c}}
\subfigure[]{
\includegraphics[scale=0.4]{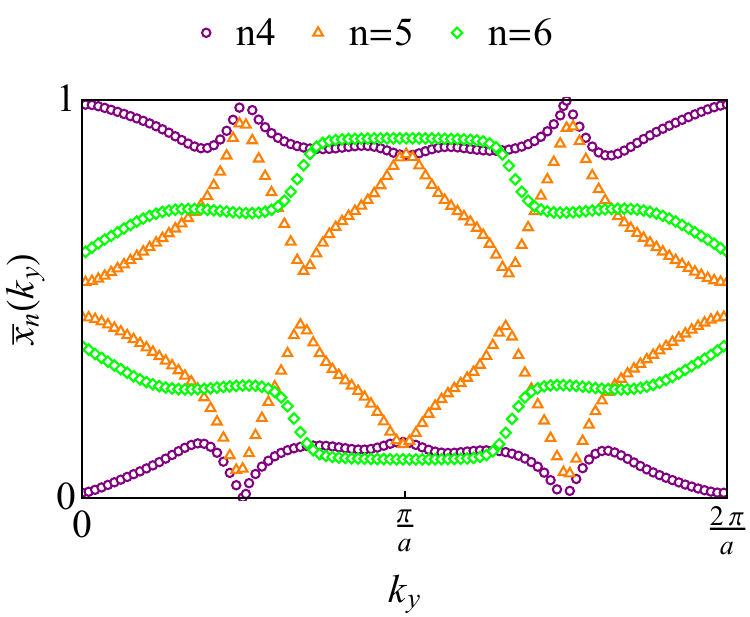}
\label{fig:7d}}
\subfigure[]{
\includegraphics[scale=0.4]{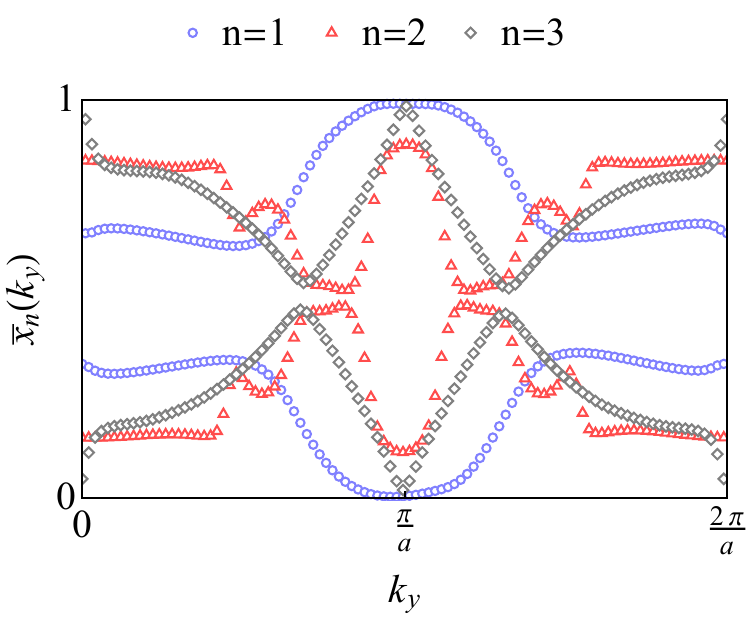}
\label{fig:7c}}
\subfigure[]{
\includegraphics[scale=0.4]{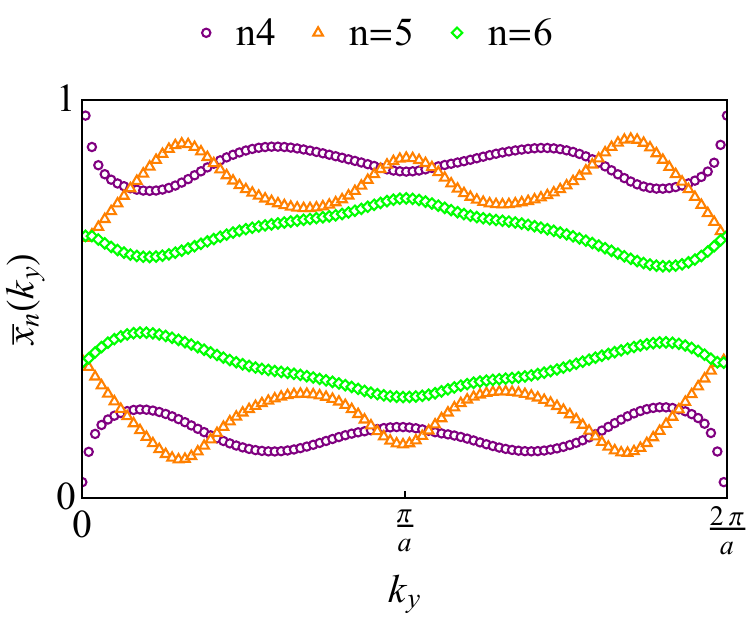}
\label{fig:7c}}
\caption[]{The spectra of Polyakov loops (Wannier centers) of individual $SU(2)$ Berry's connections of occupied bands $n=1$ through $n=6$ of Na$_{3}$Bi for various planes. The Polyakov loops are computed along $M-\Gamma-M$ or $\hat{x}$ direction and their eigenvalues can be written as $e^{\pm i \lambda_n(k_y)}$. The Wannier centers are defined as $\bar{x}_n(k_y)=\lambda_{x,n}(k_y)/(2\pi)$. (a)-(b) The Wannier spectra at $k_z=0$ mirror plane. All non-trivial bands supporting quantized, flux at the mirror plane also exhibit gapless, Wannier spectra. The summed magnitude of mirror Chern number for occupied bands is $2\pi$, hence, the $k_z=0$ mirror plane is a first-order topological insulator. (c)-(d) The gapped Wannier spectra of higher-order topological insulators with $|k_z| < k_D$. (e)-(f) The gapped Wannier spectra for topologically trivial, mirror plane $k_z=\pi/c$. The trivial and higher-order topological insulators are not distinguished by the first homotopy classification of Berry's connections. They are distinguished by the second homotopy classification or the quantized, non-Abelian Berry's flux. Therefore, the reliable topological distinction among the trivial, the first-order and the higher-order topological insulators can only be made by simultaneously computing the spectra of the Polyakov loops and the Wilson loops. }
\label{fig:Pan_6}
\end{figure*}

\begin{figure*}[t]
\centering
\subfigure[]{
\includegraphics[scale=0.16]{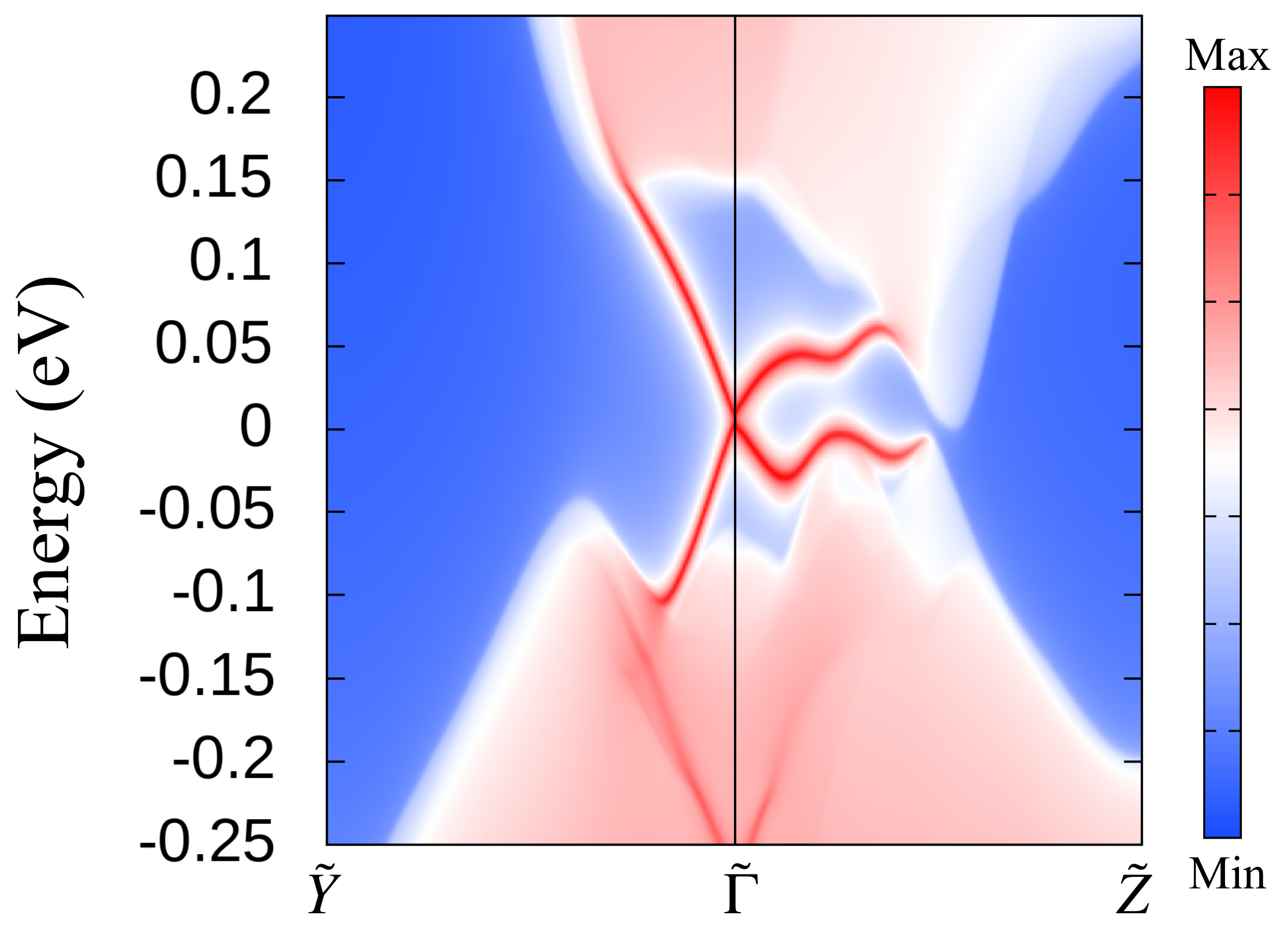}
\label{fig:CuISS}}
\subfigure[]{
\includegraphics[scale=0.21]{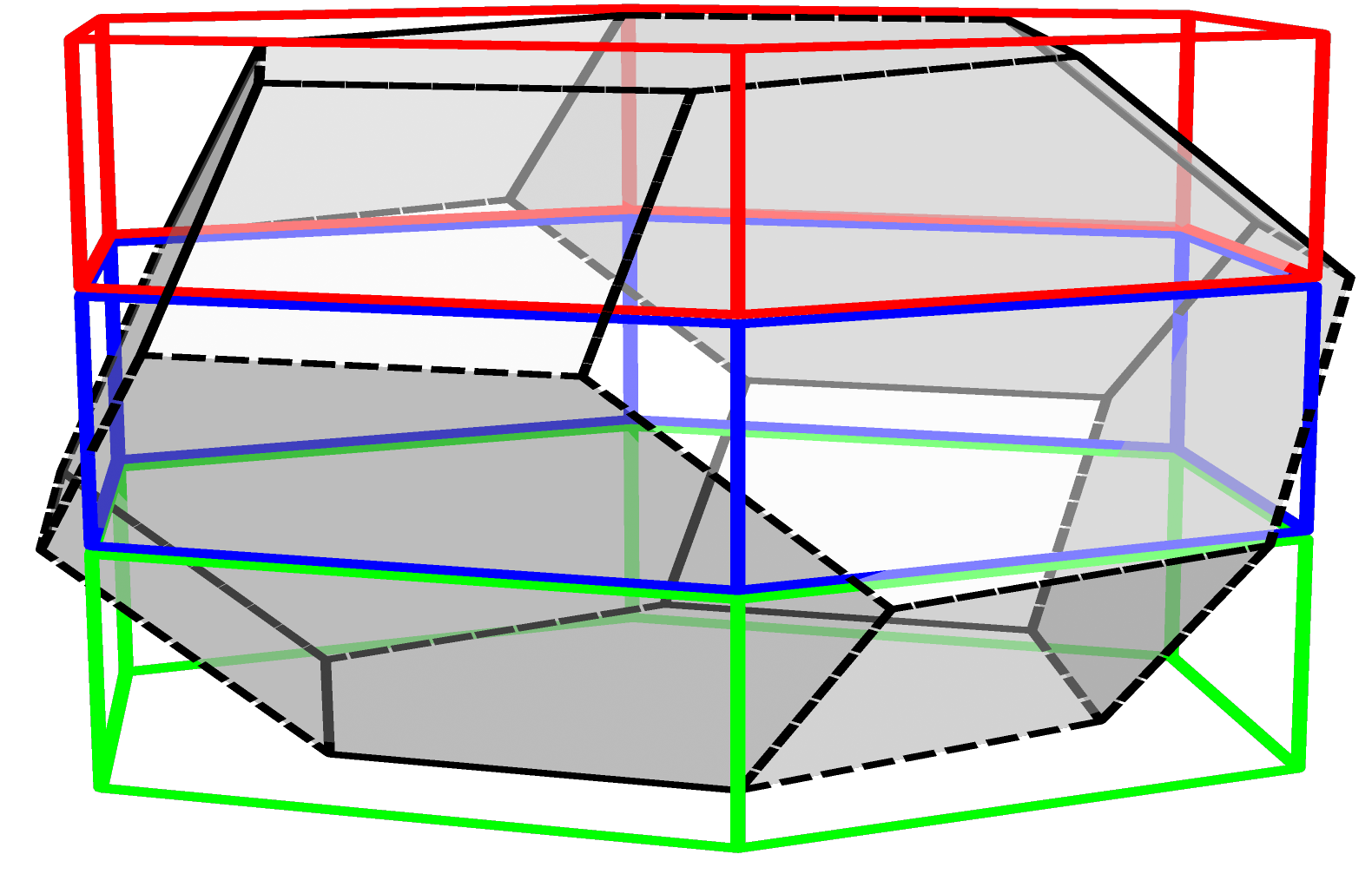}
\label{fig:PvsC}}
    \caption{(a) Spectral weight on the (100) surface of $\beta$-CuI along high-symmetry path in the vicinity of the reference energy of the Dirac point. Surface localized states are shown to begin and terminate at the projection of the Dirac points. Calculation is shown using primitive unit cell for clarity. (b) Calculation of Wilson loop requires working in the conventional unit cell. The primitive unit cell (black outline with gray shading) is shown along with three conventional unit cells colored red, green and blue, demonstrating that the volume of the primitive unit cell is three times that of the conventional unit cell, leading to a band folding effect. For details of this band folding please consult Le et. al \cite{le2018dirac}. }
\end{figure*}

\newpage

\bibliographystyle{apsrev4-1}
\nocite{apsrev41Control}
\bibliography{ref.bib}

\end{document}